\documentclass{sig-alternate-05-2015}

\usepackage{times}
\usepackage{helvet}
\usepackage{courier}
\usepackage{amsmath,amsfonts,amssymb,bbm} 
\usepackage{algorithm}  
\usepackage{algorithmic}
\usepackage{multirow}
\usepackage{extarrows}
\usepackage{graphicx}
\usepackage{xcolor}
\graphicspath{ {images/} }

\newtheorem{theorem}{Theorem}
\newtheorem{lemma}{Lemma}

\newtheorem{corollary}{Corollary}
\newtheorem{definition}{Definition}
\newtheorem{remark}{Remark}
\newtheorem{example}{Example}
\allowdisplaybreaks

\begin{document}

\setcopyright{acmcopyright}





%

\title{Security Game with Non-additive Utilities and Multiple Attacker Resources}

\numberofauthors{2} 
%
\author{
%
%
\alignauthor
Sinong Wang\\
       \affaddr{Department of ECE}\\
       \affaddr{The Ohio State University}\\
       \affaddr{Columbus, OH - 43210}\\
       \email{wang.7691@osu.edu}
\alignauthor
Ness Shroff\\
      \affaddr{Departments of ECE and CSE}\\
       \affaddr{The Ohio State University}\\
       \affaddr{Columbus, OH - 43210}\\
       \email{shroff.11@osu.edu}
}

\maketitle
\begin{abstract}
There has been significant interest in studying security games for modeling the interplay of attacks and defenses on various systems involving critical infrastructure, financial system security, political campaigns, and civil safeguarding.  However, existing security game models typically either assume additive utility functions, or that the attacker can attack only one target. Such assumptions lead to tractable analysis, but miss key inherent dependencies that exist among different targets in current complex networks. In this paper, we generalize the classical security game models to allow for non-additive utility functions. We also allow attackers to be able to attack multiple targets. We examine such a general security game from a theoretical perspective and provide a unified view. In particular, we show that each security game is equivalent to a combinatorial optimization problem over a set system $\varepsilon$, which consists of defender's pure strategy space. The key technique we use is based on the transformation, projection of a polytope, and the elipsoid method. This work settles several open questions in security game domain and significantly extends the state-of-the-art of both the polynomial solvable and NP-hard class of the security game.
\end{abstract}

%
%

%
%

%
%
\printccsdesc



\section{Introduction}

The key problem in many security domains is \emph{how to efficiently allocate limited resources to protect targets against potential threats}. For example, the government may have a limited police force to operate checkpoints and conduct random patrols. However, the adversarial aspect in security domain poses a unique challenge for allocating resources. An intelligent attacker can observe the defender's strategy and gather information to schedule more effective attacks. Therefore, the simple random strategy of ``rolling the dice''  may be exploited by the attacker, which greatly reduces the effectiveness of the strategy. 

With the development of computational game theory, such resource allocation problems can be cast in game-theoretic contexts, which provides a more sound mathematical approach to determine the optimal defense strategy. It allows the analyst to factor differential risks and values into the model, incorporate game-theoretic predictions of how the attacker would respond to the security policy, and finally determine an equilibrium strategy that cannot be exploited by adversaries to obtain a higher payoff. In the past decade, there has been an explosion of research attempting to address this approach, which has led to the development of well-known models of security games.

The classic \emph{security game} is a two-player game played between a \emph{defender} and an \emph{attacker}. The attacker chooses one target to attack; The defender allocates (randomly) limited resources, subject to various domain constraints, to protect a set of targets. The attacker (defender) will obtain the benefits (losses) for those successfully attacked targets and losses (benefits) for those defended targets.  The goal of the defender is to choose a random strategy so as to play optimally under some solution concepts such as Nash equilibrium and strong Stackelberg equilibrium. This \emph{security game} model and its game-theoretic solution is currently being used by many security agencies including US Coast Guard and Federal Air Marshals Service (FAMS)~\cite{tsai2009iris}, Transportation System Administration~\cite{brown2016one} and even in the wildlife protection~\cite{fang2013optimal}; see book by Tambe~\cite{tambe2011security} for an overview. 

\subsection{Motivation}

There exists two common limitations of the classic security game model: first, it does not consider the dependency among the different targets; second, the attacker can attack at most one target.  In particular, the payoff functions for both players are additive, i.e, the payoff of a group of targets is the sum of the payoffs of each target separately. This assumption means that the security agency measures the importance of several targets without considering the synergy among them. In practice, the attacker can simultaneously attack multiple targets and there exists some linkage structure among those targets such that attacking one target will influence the other targets. For instance, an attacker attempts to destroy the connectivity of a network and the defender aims to protect it. The strategy for both players is to choose the nodes of the network. If there are two nodes that constitute a bridge of this network, successfully attacking both of them will split the network into two parts and incur a huge damage, while attacking any one of them will have no significant effect. 

\begin{figure*}[t]
\centering
\vspace{-0.17in}
\centerline{\includegraphics[width=0.95\textwidth]{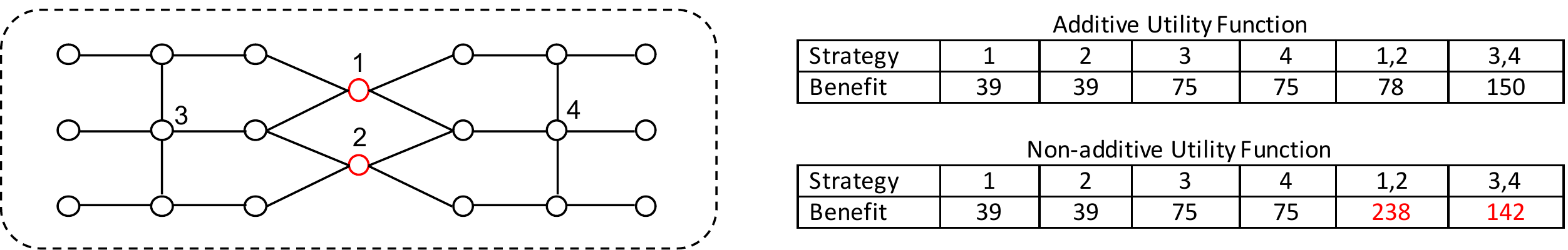}}
\caption{Example of security game in a $20-$nodes network.}
\vspace{-0.1in}
\label{fig:example}
\end{figure*}

\begin{example}
As shown in Fig.~\ref{fig:example}, we have a $20-$node network. It is clearly that nodes $1, 2, 3$ and $4$ are the critical battlefields in this network. Suppose that the attacker's and defender's strategies are $\{1\}, \{2\}, \{3\}, \{1, 2\}$ or $\{3,4\}$, where $\{v\}$ denotes the index of the nodes. We adopt the network value function proposed by~\cite{gueye2012towards} as the security measure for different nodes, which calculates the importance of a group of nodes via subtracting the value of the network by removing these nodes from the value of the original network\footnote{Compared with traditional measures such as node degree or betweenness centrality, the network value provides a more accurate description of the importance of different nodes. We can use the other kinds of network values instead of the quadratic form.}. For example, if we adopt the network value as a function $f(\{n_i\})=\sum_{i} n_i^2$, where $n_i$ is number of nodes in the $i$th connected component, the value of the original network is $20^2=400$. After removing node $3$, the network will be divided into two components: one $18-$node network and one isolated node, the network value is reduced to $18^2+1^2=325$. Thus the benefit of node 3 is equal to the decrement $400-325=75$. Similarly, we can get the benefits of other nodes as illustrated in the bottom table of Fig~\ref{fig:example}. In traditional security game models~\cite{korzhyk2011security}, they assume that the benefit of strategy $\{1,2\}$ and $\{3, 4\}$ is equal to $39+39=78$ and $75+75=150$. The mixed strategy equilibrium\footnote{In this example, we adopt the zero-sum security game model.} under this case is that defender choose nodes $1,2$ with probability $0.34$ and nodes $3,4$ with probability $0.66$. Instead, if we adopt the true value of nodes $\{1,2\}$ and $\{3,4\}$ (as illustrated in red of bottom table), the equilibria is that the defender chooses nodes $1,2$ with probability $0.63$ and nodes $3,4$ with probability $0.37$. From the point view of the network, the second one provides a more reliable strategy.
\end{example}

As can be seen in Example 1, the traditional models that ignore the inherent synergy effect between the targets are limiting and could lead to catastrophic consequences. A comprehensive analysis regarding the effect of dependent targets can be found in~\cite{vorobeychik2015securing}. 

\subsection{Related Works}

The nature of resource allocation in \emph{security games} often results in exponentially many pure strategies for the defender, such that the defender's optimal mixed strategy is hard to solve. In the past decades, there have been numerous algorithms developed for various extensions of the classical \emph{security game} model discussed earlier.

One line of research focuses on designing an efficient algorithm to solve such a game.  \cite{kiekintveld2009computing} proposes a compact representation technique, in which the security game can be equivalently represented by a polynomial-sized mixed-integer linear programming (MILP) problem. The issue in~\cite{kiekintveld2009computing} is that they only determine the optimal solution of the compact game instead of the optimal defender's mixed strategy. To solve this problem, \cite{korzhyk2010complexity} introduces the Birkhoff-von Neumann theorem and show that the defender's mixed strategy can be recovered under a specific condition. \cite{jain2011double} proposes a double-oracle algorithm to exactly solve the security game with exponential large representation, which can be regarded as a generalization of traditional column generation technique in solving the large-scale linear programming problem. There are also other works such as the Bayesian security game~\cite{paruchuri2008playing}, the security game with quantal response~\cite{an2012protect} and the security game with uncertain attacker behavior~\cite{balcan2015commitment}, etc.

Another line of research focuses on examing the complexity of the security game. ~\cite{korzhyk2010complexity} adopts the previous compact representation framework~\cite{kiekintveld2009computing} and shows that the compactly represented security game is polynomial solvable in some cases while NP-hard in other cases. \cite{xu2014solving} shows that the spatial and temporal security game is generally NP-hard. \cite{korzhyk2011stackelberg} provides an interesting result that, if the attacker has a single resource, the strong Stackelberg equilibrium is also a Nash equilibrium, which resolves the leader's dilemma; if the attacker has multiple resources, this property does not hold. \cite{bhattacharya2011approximation} proves that the general security game with costly resources\footnote{Here ``costly resources'' means that the defender's resource are obtained at some costs and the defender has a budget to allocate his resources.} is NP-hard and proposes an approximation algorithm. 

The earlier mentioned works studying the complexity of the security game focus on the game with a single attacker resource. However, none of these works provide a systematic understanding of complexity properties or provide an efficient algorithm for the security game when attacker has multiple resources and utility functions are non-additive. In~\cite{korzhyk2011security}, the authors extend the classic security game model to the scenario of multiple attacker resources. They design a $6-$states transition algorithm to exactly compute the Nash equilibrium in polynomial time. Such an algorithm is complicated and restricted to the case that the defender's resource is homogenous, i.e., the defender can protect any subset of targets with a cardinality constraint. In the practical scenario such as FAMS, the defender's resources may be heterogenous and solving such a scenario is still an open question in the security game domain.

In work~\cite{wang2016non}, the authors propose to investigate the security game with non-additive utilities. However, they assume that the  security game is zero-sum, the defender's resources are homogenous and only one one group of utilities (benefit function). The recent work~\cite{xu2016mysteries} provides a unified framework of the classical security game model with a single attacker's resource. He shows that solving the security game is equivalent to solving a linear optimization problem over a set system. Using the flexibility of such a set system, their framework can encode most previous security games. For example, if the set system is a uniform matroid, it recovers the result of LAX checkpoint placement problem~\cite{pita2008deployed}; if the set system represents a coverage problem, it recovers most results in~\cite{korzhyk2010complexity}, i.e., the polynomial solvability of $2-$weighted coverage implies the polynomial solvability of equilibrium computation; if the set system represents the problem of independent set, it recovers the NP-hardness result in~\cite{brown2016one}. However, they leave open the question: what is the complexity of the security game model when attacker has multiple resources and utility functions are non-additive. 

\begin{table*}[htb]%
\centering
\label{tab:result}
\caption{Solution Status in the Security Game\label{tab:one}}{%
\begin{tabular}{|c|c|c|c|c|}
\hline
\multicolumn{2}{|c|}{Cases} & Single attacker resource & \multicolumn{2}{|c|}{Multiple attacker resource}\\
\cline{4-5}
\multicolumn{2}{|c|}{} &  & Homogenous resource & Heterogenous resource\\
\hline
\multirow{2}{*}{Additive utility function} & Zero-sum & SSE,NE~\cite{kiekintveld2009computing,korzhyk2011stackelberg,xu2016mysteries,letchford2013solving} & SSE,NE~\cite{korzhyk2011security} &  \textbf{SSE, NE Theorem~\ref{thm:organddop}, Lemma~\ref{lm:add}}\\
\cline{2-5}
& Non-zero-sum & SSE~\cite{letchford2013solving}, NE~\cite{korzhyk2011stackelberg,xu2016mysteries}& NE~\cite{korzhyk2011security} & \textbf{SSE Theorem~\ref{thm:sseanddop}},\textbf{NE Theorem~\ref{thm:dopandnonzero}}\\
\hline
\multirow{2}{*}{Non-additive utility function} & Zero-sum & Same as above & \multicolumn{2}{|c|}{\textbf{NE, SSE Theorem~\ref{thm:organddop}, Theorem~\ref{thm:dop}}}\\
\cline{2-5}
& Non-zero-sum & Same as above &\multicolumn{2}{|c|}{\textbf{ SSE Theorem~\ref{thm:sseanddop},Theorem~\ref{thm:dop}}} \\
\hline
\end{tabular}}
\vspace{-0.1in}
\end{table*}%

\subsection{Our Results}

In this paper, we study the classic \emph{security game} model when attacker has  multiple resources and utility functions are non-additive. More specifically, we wonder how the following questions that are well understood in the case of single attacker resource and additive utility functions can be addressed in this general case.
\begin{itemize}
  \item How to compactly represent the security game with multiple attacker resources and non-additive utility functions?
  \item How to efficiently solve such a compactly represented game?
  \item What is the complexity of the security game when we consider non-additive utility functions and allow the attackers to attack multiple attacker resources?
\end{itemize}

To answer these questions, we provide the following contributions: (1) we first propose a polytope transformation and projection framework to equivalently and compactly represent the zero-sum and non-additive security game with only poly($n$) variables; (2) We  prove that the problem of determining the Nash equilibrium of zero-sum and non-additive security game and the problem of optimizing a \emph{Pseudo-Boolean function} over a set system $\varepsilon$ can be reduced to each other in polynomial time. The main technique we use is to exploit the geometric structure of the low-dimensional polytope to construct a polynomial time vertex mapping algorithm. (3) We then apply our framework to the non-zero-sum and non-additive security game, and further obtain a similar result that determining the strong Stackelberg equilibrium and the above combinatorial optimization problem is equivalent. (4) Finally, we examine the Nash equilibrium in the non-zero-sum but additive security game. We prove that determining the Nash equilibrium can be reduced to the linear optimization over a set system $\varepsilon$. The basic technique is to add another polytope transformation step in our previous framework, and show that determining the Nash equilibrium of non-zero-sum game can be reduced to a polynomial dimensional saddle point problem. The main results and comparison is summarized in TABLE~\ref{tab:result}.

These results demonstrate that the security game with non-additive utility function and multiple attacker resource is essentially a combinatorial problem, and provide a systematic framework to transform the game-theoretical problem to the problem of combinatorial algorithm design. Further, our results not only answers the questions proposed in the security game domain~\cite{korzhyk2011security,xu2016mysteries}, but also extends significantly both the polynomial solvable and NP-hard class:
\begin{itemize}
  \item The previous result in~\cite{xu2016mysteries} is dependent on the description length of the set system. One special case of our result  can recover and strengthen the main results in~\cite{xu2016mysteries}, which is independent of the description length of the set system. It also implies that we also recover and strengthen the results in most security game papers.
  \item The polynomial time solvability in~\cite{korzhyk2011security} corresponding to that the set system $\varepsilon$, is a uniform matroid, which can be easily solved by summing the first $k$ largest elements. In the scenario of the heterogenous resource, we can extend several polynomial solvable classes.
  \item Solve the security game occurred in the tree network or a sparse network. In these cases. although the utility functions are non-additive, we can show such a problem can be reduced to a polynomial solvable oracle such as a sub-modular minimization problem.
\end{itemize}

The detailed discussions can be seen in Section 6. The rest of the paper is organized as follows. In Section 2, we introduce the security game model with multiple attacker resources and non-additive utility functions, and the preliminaries regarding some classical results in combinatorial optimization. In Section 3, we present our framework of polytope transformation and projection to compactly represent the zero-sum security game. In Section 4, we present the reduction between the zero-sum security game and the combinatorial optimization problem. In Section 5, we further generalize our framework to the non-zero-sum security game and present the main results. Finally, we provides numerous applications of our theoretical framework in Section 6. We conclude our work in Section 7. Due to the space limitation, all of our technical proofs are provided in the Appendix.


\section{Model and Preliminary}

In the following, we first define the security game with non-additive utility functions and multiple attacker resources as a two-player normal-form non-zero-sum game. Then we present several classic results in combinatorial optimization.

\subsection{Problem Description}

The model is similar to the classic security game~\cite{kiekintveld2009computing}, and the only difference is that we consider multiple attacker resources and non-additive utility functions.

\textbf{Players} and \textbf{targets:} The security game contains two players (a \emph{defender} and an \emph{attacker}), and $n$ targets. We use $[n]\triangleq \{1,2,\ldots,n\}$ to denote the set of these targets. 

\textbf{Strategies and index function:} The \emph{pure} strategy for each player is the subset of targets and all the pure strategies for each player constitute a collection of subsets of $[n]$.  We assume that the attacker can attack at most $c$ targets, where $c>1$ is a constant\footnote{Later, in Section 5, we will relax this constant assumption.}.  The attacker's pure strategy space is a uniform matroid $\mathcal{A}=\{A\subseteq[n]| |A|\leq c\}$ and the number of attacker's pure strategies is $N_a\triangleq|\mathcal{A}|$. Similarly, we use $\mathcal{D}\in 2^{[n]}$ to denote the defender's pure strategy space and $N_d\triangleq|\mathcal{D}|$.  Note that there exists some resource allocation constraints in practice and such that $\mathcal{D}$ is not always a uniform matroid. For example, if the defender has a budget and its resource are obtained at some costs, in which the costs are heterogenous. In this case, the defender's feasible pure strategy is all the possible combinations of the targets with total cost less than the budget.

Suppose that the order of the pure strategy of the attacker is given by index function $\sigma(\cdot)$, which is a one-one mapping: $2^{[n]}\rightarrow \{1,2,\cdots,2^n\}$. Then, we define the following index function $\mu(\cdot)$ for the pure strategy of the defender as: $\mu(U)=\sigma(U^c)$ for any $U\in2^{[n]}$. For simplicity, the index function $\sigma(\cdot)$ and $\mu(\cdot)$ are defined over all subsets of $[n]$. The reason behind this definition of the index function is to simplify the representation of most theoretical results. For example, if $n=2$, $\mathcal{A}=\mathcal{D}=2^{\{1,2\}}$, and the order of the attacker's pure strategy is $\sigma(\{1,2\})=1, \sigma(\{2\})=2, \sigma(\{1\})=3$ and $\sigma(\{\emptyset\})=4$, then the order for defender's pure strategy is $\mu(\{\emptyset\})=1, \mu(\{1\})=2, \mu(\{2\})=3$ and $\mu(\{1,2\})=4$.

The \emph{mixed} strategy is the probability distribution over the pure strategy space, which is employed when the player chooses its strategy based on some random experiment. Specifically, if the attacker chooses $\mathbf{p}$ as its mixed strategy, the probability that the strategy $A$ is chosen is $\mathbf{p}_{\sigma(A)}$. The set of all the mixed strategies of attacker and defender can be represented as the simplex $\Delta_{N_a}$ and $\Delta_{N_d}$, where
\begin{equation*}
  \Delta_{N_a}=\{\mathbf{p}\in\mathbb{R}^{N_a}_{+}|\sum_{A\in \mathcal{A}}\mathbf{p}_{\sigma(A)}=1\}.
\end{equation*}
Similar definition holds for $\Delta_{N_d}$.

\begin{figure}[t]
\vspace{-0.0in}
\centerline{\includegraphics[width=0.43\textwidth]{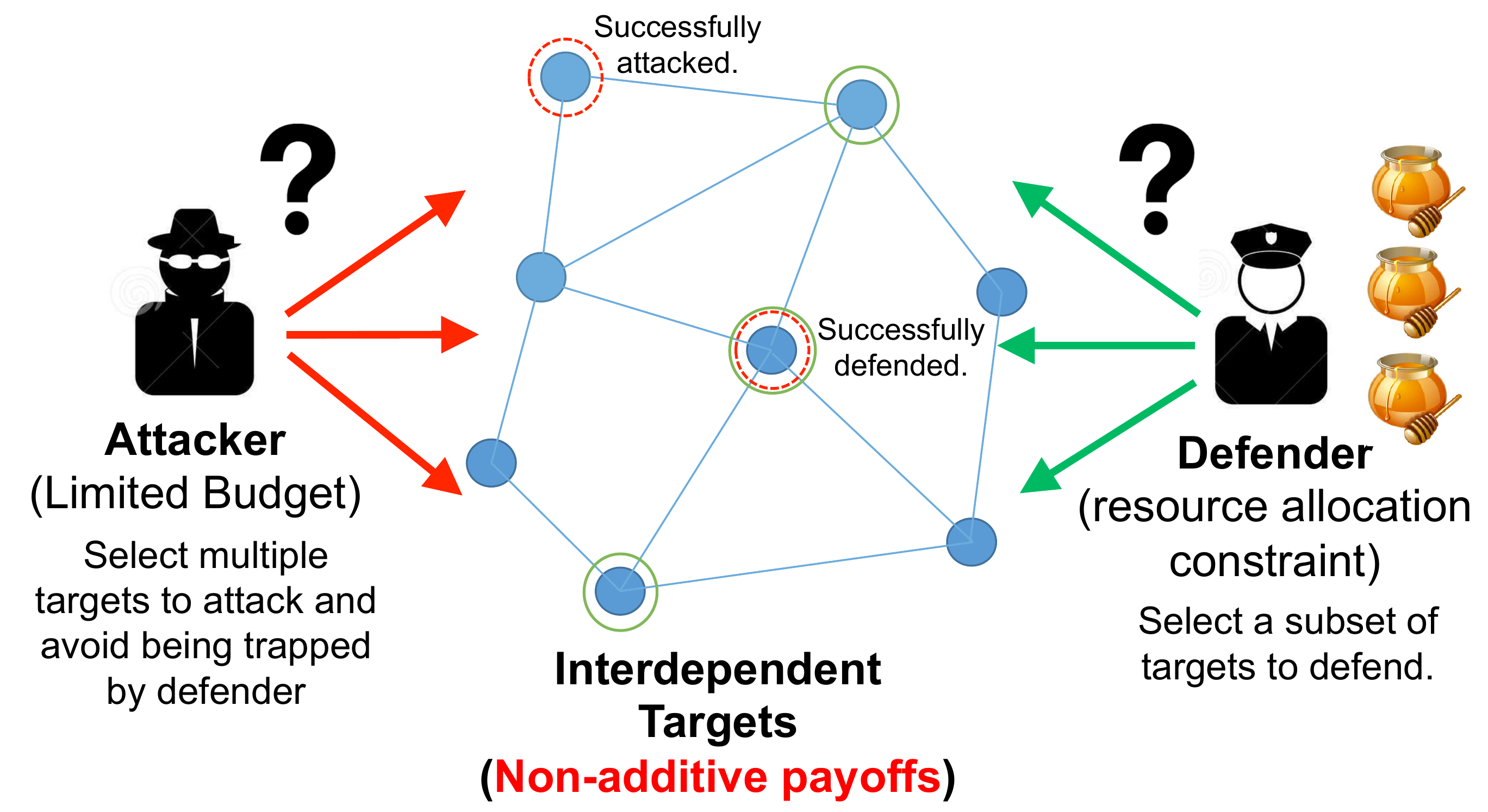}}
\vspace{-.1in}
\caption{Security game with non-additive utility functions and multiple attacker resources.}
\vspace{-0.1in}
\label{fig:model}
\end{figure}

\textbf{Payoff Structure:}  
The benefits and losses are represented by utility functions as follows. Let set function $B_{a}(\cdot):\mathcal{A}\rightarrow \mathbb{R}$ and $B_{d}(\cdot):\mathcal{D}\rightarrow \mathbb{R}$ be the attacker's and defender's benefit functions, and the set function $L_a(\cdot): \mathcal{A}\rightarrow \mathbb{R}$ and $L_d(\cdot): \mathcal{D}\rightarrow \mathbb{R}$ be the corresponding loss functions. The standard assumption is that the benefit is always larger than the loss: $B_{a}(A)>L_{a}(A)$ and $B_{d}(A)>L_{d}(A)$ for all $A\in\mathcal{A}$. If the attacker and defender choose strategy $A\in\mathcal{A}$ and $D\in\mathcal{D}$, the attacker's and defender's payoff is given by $B_a(A\backslash D)+L_{a}(A\cap D)$ and $B_d(A\cap D)+L_{d}(A\backslash D)$, respectively\footnote{$A\backslash D$ is the standard set difference, defined by $A\backslash D=\{x|x\in A, x\notin D\}$ and is equal to $A\cap D^c$, where $D^c$ is the complementary set of subset $D$.}. A security game is zero-sum if $B_a(A\backslash D)+L_{d}(A\backslash D)=0$ and $B_d(A\cap D)+L_{a}(A\cap D)=0$ for any $A$ and $D$, which means that one player's benefit is indeed the loss of the other player.

\textbf{Bilinear-form:} Based on the above payoff structure,  we can define the benefit matrices of attacker and defender: $\mathbf{B}^a$ and $\mathbf{B}^d\in\mathbb{R}^{N_a\times N_d}$: $\forall A\in\mathcal{A},D\in\mathcal{D}$,
\begin{equation*}
	\mathbf{B}^a_{\sigma(A),\mu(D)}=B_a(A\backslash D), \mathbf{B}^d_{\sigma(A),\mu(D)}=B_d(A\cap D),
\end{equation*}
and the loss matrices: $\mathbf{L}^a$ and $\mathbf{L}^d\in\mathbb{R}^{N_a\times N_d}$: $\forall A\in\mathcal{A},D\in\mathcal{D}$,
\begin{equation*}
	\mathbf{L}^a_{\sigma(A),\mu(D)}=L_a(A\cap D), \mathbf{L}^d_{\sigma(A),\mu(D)}=L_d(A\backslash D),
\end{equation*}
Let $\mathbf{M}^a$ and $\mathbf{M}^d$ be the attacker's and defender's payoff matrices. It is clear that $\mathbf{M}^a=\mathbf{B}^a+\mathbf{L}^a$ and $\mathbf{M}^d=\mathbf{B}^d+\mathbf{L}^d$. Then the expected payoffs for the attacker and defender is given by following bilinear form, when they play the mixed strategy $\mathbf{p}\in\Delta_{N_a}$ and $\mathbf{q}\in\Delta_{N_d}$, by 
\begin{align*}
U_a(\mathbf{p},\mathbf{q})=\mathbf{p}^T\mathbf{M}^a\mathbf{q} \quad\text{and}\quad U_d(\mathbf{p},\mathbf{q})=\mathbf{p}^T\mathbf{M}^d\mathbf{q}.
\end{align*}

\textbf{Solution Concepts:} If both players move simultaneously, the standard solution concept is the \emph{Nash equilibrium (NE)}, in which no single player can obtain a higher payoff by deviating unilaterally from this strategy. A pair of mixed strategies $(\mathbf{p^*},\mathbf{q^*})$ forms a \emph{NE} if and only if they satisfy the following: $\forall \mathbf{p}\in\Delta_{N_a}, \mathbf{q}\in\Delta_{N_d}$,
\begin{align*}
   &U_{d}(\mathbf{p^*},\mathbf{q^*})\geq U_{d}(\mathbf{p^*},\mathbf{q})\text{ and }U_{a}(\mathbf{p^*},\mathbf{q^*})\geq U_{a}(\mathbf{p},\mathbf{q^*}).
\end{align*}

In some application domain, the defender can build fortifications before the attack and is thus in the leader's position from the point view of the game, and able to move first. In this case, the \emph{strong Stackelberg equilibrium (SSE)} serves as a more appropriate solution concept~\cite{von1934marktform,von2004leadership}, where the defender commits to a mixed strategy; the attacker observes this strategy and comes up with its best responses. Formally, let $C(\mathbf{q})=\arg\max_{\mathbf{p}\in\Delta_{N_a}}U^{a}(\mathbf{p},\mathbf{q})$ denote the attacker's best response to defender's mixed strategy $\mathbf{q}$. A pair of mixed strategies $(\mathbf{p^*},\mathbf{q^*})$ is a SSE, if and only if,
\begin{align*}
   &\mathbf{q^*}=\arg\max\limits_{\mathbf{q}\in\Delta_{N_d}}U_d(C(\mathbf{q}), \mathbf{q}) \text{ and } \mathbf{p^*}=C(\mathbf{q^*}).
\end{align*}
Our goal is to compute the defender's Nash equilibrium strategies and strong Stackelberg equilibrium strategies and we call it as \textbf{equilibrium computation problem}. The following definitions are often used in our theoretical development.
\begin{definition}\label{def:com}
\emph{(Common utility)}
 	The common utility is defined as the M$\ddot{o}$bius transformation~\cite{kennes1990computational,fomin2010exact} of the benefit and loss function $B_a(U)$ and $L_a(U)$ for all $U\in 2^{[n]}$,
	\begin{align*}
		&B^c_a(U)=\sum\limits_{V\subseteq U}(-1)^{|U\backslash V|}B_a(V),\text{ } L^c_a(U)=\sum\limits_{V\subseteq U}(-1)^{|U\backslash V|}L_a(V).
	\end{align*}
\end{definition}
Similar definitions hold for defender's benefit and loss function: $B_d(\cdot)$, $L_{d}(\cdot)$ and their common utilities: $B^c_d(\cdot)$, $L^{c}_d(\cdot)$.

\begin{definition}\label{def:supset}
\emph{(Support set)} The support set of the security game is defined as
\begin{equation}
  S=\{A\in \mathcal{A}|B^c_a(A)\text{ or }B^c_d(A)\text{ or }L^c_a(A)\text{ or }L^c_d(A)\neq 0 \}.
\end{equation}
and the support index set $\sigma(S)=\{\sigma(A)|A\in S\}$.
\end{definition}

\begin{definition}
\emph{(Projection operator)} The projection operator $\pi_S: \mathbb{R}^N\rightarrow\mathbb{R}^{|S|}$ is
\begin{equation}
	\pi_S((\mathbf{x}_1,\mathbf{x}_2,\ldots,\mathbf{x}_N))=(\ldots,\mathbf{x}_{i},\ldots)_{i\in \sigma(S)},
\end{equation} 
and projection of polytope: $\Pi_{S}(\Delta_N)\triangleq \{\pi_S(\mathbf{x})|\mathbf{x}\in\Delta_N\}$. 
\end{definition}
The following definition of the set system $\varepsilon$ is a binary representation of the defender's pure strategy space $\mathcal{D}$.
\begin{definition}\label{def:setsystem}
\emph{(Set System)} The set system $\varepsilon\triangleq\{\mathbf{x}\in\{0,1\}^n|\mathbf{x}_i=\mathbbm{1}\{i\in D\}, \forall 1\leq i\leq n, \forall D\in\mathcal{D}\}$.
\end{definition}

\subsection{Preliminaries}

Let $H$ be a non-empty convex polytope in $\mathbb{R}^n$. Given a vector $\mathbf{w}\in\mathbb{R}^n$, one wants to find a solution to $\max_{\mathbf{x}\in H}\mathbf{w}^T\mathbf{x}$. By ``linear optimization over $H$'', we mean solving the problem $\max_{\mathbf{x}\in H}\mathbf{w}^T\mathbf{x}$ for any $\mathbf{w}\in \mathbb{R}^n$. A separation problem for $H$ is that, given a vector $\mathbf{x}\in\mathbb{R}^n$, decide if $\mathbf{x}\in H$, and if not, find a hyperplane which separates $\mathbf{x}$ from $H$. The following results are due to Gr$\ddot{o}$tschel, Lovasz and Schrijver~\cite{grotschel1981ellipsoid}.

\begin{theorem}\label{thm:equioptandsep}
\emph{(Separation and optimization)}  Let $H\in\mathbb{R}^n$ be a convex polytope. There is a \emph{poly}($n$) time algorithm to solve the linear optimization problem over $H$ if and only of there is a \emph{poly}($n$) time algorithm to solve the separation problem for $H$.
\end{theorem}

\begin{theorem}\label{thm:convdecomp}
\emph{(Separation and convex decomposition)} Let $H\in\mathbb{R}^n$ be a convex polytope. If there is a \emph{poly}($n$) time algorithm to solve the separation problem for $H$, then there is a \emph{poly}($n$) time algorithm that, given any $\mathbf{x}\in H$, yields $(n+1)$ vertices $\mathbf{v}^1,\ldots,\mathbf{v}^{n+1}\in H$ and convex coefficients $\lambda_1,\ldots,\lambda_{n+1}$ such that $\mathbf{x}=\sum_{i=1}^{n+1}\lambda_i\mathbf{v}^i$.
\end{theorem}

The following result is the generalization of the von Neumann's minimax theorem, which provides a condition when we can use the minimax or maximin formulation to solve a saddle point problem~\cite{sion1958general}.

\begin{theorem}
\emph{(Sion's minimax theorem)}  Let $X$ be be a compact convex subset of a linear topological space and $Y$ is a convex subset of a linear topological space. If $f$ is a real-valued function on $X\times Y$ with 
\begin{enumerate}
  \item $f(x,\cdot)$ is lower semicontinuous and quasi-convex on $y,\forall x\in X$ and 
  \item $f(\cdot,y)$ is upper semicontinuous and quasi-concave on $x,\forall y\in Y,$
\end{enumerate}
then we have
\begin{equation}
 \min\limits_{y\in Y}\max\limits_{x\in X} f(x,y)=\max\limits_{x\in X}\min\limits_{y\in Y} f(x,y).
\end{equation}
\end{theorem}

\section{The Compact Representation \\for Zero-sum Security Game}

The Nash equilibrium is equivalent to the strong Stackelberg equilibrium in the zero-sum game. Therefore, we only focus on the computation of Nash equilibrium. Invoking the result in the von Neumann's minimax theorem, computing the NE of zero-sum game can be formulated as the following minimax problem,
\begin{align}
	&\min\limits_{\mathbf{q}\in \Delta_{N_d}}\max\limits_{\mathbf{p}\in\Delta_{N_a}}U_a(\mathbf{p},\mathbf{q})=\mathbf{p}^T\left(\mathbf{B}^{a}+\mathbf{L}^{a}\right)\mathbf{q}\label{eq:orgmodel}.
\end{align} 
Although it can be cast into a linear programing problem, such an optimization model has $\Omega(n^k)$ variables, which is exponential in $n$ in the worst case, i.e., the defender can protect any subsets of targets. The goal of this section is to develop a technique to compactly and equivalently represent the zero-sum and non-additive security game with only $\text{poly}(n)$ variables. To convey our idea more easily, we begin with an example.

\subsection{Motivating Example}

We first use gauss elimination on matrices $\mathbf{B}^{a}$ and $\mathbf{L}^{a}$ to transform them into row canonical form, which is to left and right multiply such matrices by elementary matrices $\mathbf{E}_1, \mathbf{E}_2\in\mathbb{R}^{N_a\times N_a}$ and $\mathbf{F}_1, \mathbf{F}_2\in\mathbf{R}^{N_d\times N_d}$.
\begin{align}
	&\min\limits_{\mathbf{q}\in \Delta_{N_d}}\max\limits_{\mathbf{p}\in\Delta_{N_a}}\mathbf{p}^T\left(\mathbf{B}^{a}+\mathbf{L}^{a}\right)\mathbf{q}\notag\\
	&=\min\limits_{\mathbf{q}\in \Delta_{N_d}}\max\limits_{\mathbf{p}\in\Delta_{N_a}}\mathbf{p}^T\mathbf{E_1}\mathbf{E}^{-1}_1\mathbf{B}^{a}\mathbf{F}^{-1}_1\mathbf{F_1}\mathbf{q}+\mathbf{p}^T\mathbf{E}_2\mathbf{E}^{-1}_2\mathbf{L}^{a}\mathbf{F}_2^{-1}\mathbf{F}_2\mathbf{q}\notag\\
	&=\min\limits_{\mathbf{q}\in \Delta_{N_d}}\max\limits_{\mathbf{p}\in\Delta_{N_a}}\mathbf{p}^T\mathbf{E}_1\begin{bmatrix}
\mathbf{B}^{a}_{r} & \mathbf{0}\\ 
\mathbf{0} & \mathbf{0}
\end{bmatrix}\mathbf{F}_1\mathbf{q}+\mathbf{p}^T\mathbf{E}_2\begin{bmatrix}
\mathbf{L}^{a}_{s} & \mathbf{0}\\ 
\mathbf{0} & \mathbf{0}
\end{bmatrix}\mathbf{F}_2\mathbf{q}.\notag
\end{align}
where $r$ and $s$ are the rank of matrices $\mathbf{B}^{a}$, $\mathbf{L}^{a}$,  and $\mathbf{B}^{a}_{r}$, $\mathbf{L}^{a}_{s}$ are the corresponding non-zero blocks of their row canonical form. If we define the affine transformation: $f_1(\mathbf{p})=\left(\mathbf{p}^T\mathbf{E}_1\right)^T$, $f_2(\mathbf{p})=\left(\mathbf{p}^T\mathbf{E}_2\right)^T$, $ g_1(\mathbf{q})=\mathbf{F}_1\mathbf{q}$ and $g_2(\mathbf{q})=\mathbf{F}_2\mathbf{q}$. Let\footnote{The notation $(\cdot,\cdot)$ denotes the concatenation operator of vector.}
\begin{align*}
  &\Delta^{a}_{N_a}=\{(f_1(\mathbf{p}),f_2(\mathbf{p}))| \mathbf{p} \in \Delta_{N_a}\},\\
  &\Delta^{d}_{N_d}=\{(g_1(\mathbf{q}),g_2(\mathbf{q}))|\mathbf{q} \in \Delta_{N_d}\}.
\end{align*}
we can obtain the following equivalent optimization problem,
\begin{align*}
	&\min\limits_{(\bar{\mathbf{q}}_1,\bar{\mathbf{q}}_2)\in\Delta^{d}_{N_d}}\max\limits_{(\bar{\mathbf{p}}_1,\bar{\mathbf{p}}_2)\in\Delta^{a}_{N_a}}\bar{\mathbf{p}}_1^T\begin{bmatrix}
\mathbf{B}^{a}_{r} & \mathbf{0}\\ 
\mathbf{0} & \mathbf{0}
\end{bmatrix}\bar{\mathbf{q}}_1+\bar{\mathbf{p}}_2^T\begin{bmatrix}
\mathbf{L}^{a}_{s} & \mathbf{0}\\ 
\mathbf{0} & \mathbf{0}
\end{bmatrix}\bar{\mathbf{q}}_2.
\end{align*}

Moreover, considering the fact that only the first $r$ elements in vector $\bar{\mathbf{p}}_1$ and $\bar{\mathbf{q}}_1$  and first $s$ elements in $\bar{\mathbf{p}}_2$ and $\bar{\mathbf{q}}_2$ have the non-zero coefficients in the above optimization model, we can further simplify the above optimization problem as 
\begin{align}
	&\min\limits_{(\bar{\mathbf{q}}_1,\bar{\mathbf{q}}_2)\in H_d}\max\limits_{(\bar{\mathbf{p}}_1,\bar{\mathbf{p}}_2)\in H_a}\bar{\mathbf{p}}_1^T\mathbf{B}^{a}_{r}\bar{\mathbf{q}}_1+\bar{\mathbf{p}}_2^T\mathbf{L}^{a}_{s}\bar{\mathbf{q}}_2\label{eq:excptmodel},
\end{align}
where the $H_a$ and $H_d$ is obtained by projecting the polytope $\Delta^{a}_{N_a}$ and $\Delta^{d}_{N_d}$ to those coordinates belonging to the non-zero blocks.

The basic observation in the above example is that the number of variables in the optimization model  (\ref{eq:excptmodel}) is equal to the sum of rank $r+s$ of payoff matrices. Based on the rank inequality that the rank of a matrix is less than its dimension, we have that $r,s \leq \min\{N_a,N_d\}$.  Since the number of attacker's pure strategies is $N_a=O(n^c)=\text{poly}(n)$. Therefore,  there exists at most poly($n$) variables in the optimization model (\ref{eq:excptmodel}).

\subsection{Formal Description}

The above illustrative derivation provides a possible path to compactly represent the game. However, there exists a significant technical challenge: the elementary matrices $\mathbf{F}_1$, $\mathbf{F}_2$ and their inverse matrices may have exponential size due to exponential large defender's pure strategy space. Hence, the key question is whether we can \textbf{find both these elementary matrices} efficiently? To tackle this problem, we first show that payoff matrices $\mathbf{B}^{a}$ and $\mathbf{L}^{a}$ can be decomposed as the product of the several simple matrices. The following technical lemma is critical in our decomposition.
\begin{lemma}\label{lm:com}
 \emph{(Utility and common utility)} For all $U\in 2^{[n]}$, the benefit and loss functions satisfy
	\begin{align*}
		&B_a(U)=\sum\limits_{V\subseteq U}B^c_a(V)\text{ and } L_a(U)=\sum\limits_{V\subseteq U}L^c_a(V).
	\end{align*}
\end{lemma}

The lemma~\ref{lm:com} provides a path to recover the utility functions from the common utility. Indeed, it is named as the zeta transformation~\cite{fomin2010exact}, which is the inverse transformation of the M$\ddot{o}$bius transformation given in Definition~\ref{def:com}. Suppose that $\mathbf{B}^{A}$ and $ \mathbf{L}^{A}$ represent the benefit and loss matrix for attacker when $\mathcal{A}=\mathcal{D}=2^{[n]}$, and $\mathbf{B}^{a},  \mathbf{L}^{a}$ can be regarded as the sub-matrix of  $\mathbf{B}^{A}, \mathbf{L}^{A}$. The following  technical lemma presents the decomposable property of payoff matrices $\mathbf{B}^{A}$ and $ \mathbf{L}^{A}$.
\begin{lemma}\label{lm:decomplete}
\emph{(Decomposition of complete payoff matrix)} If the attacker's and defender's pure strategy space $\mathcal{A}=\mathcal{D}=2^{[n]}$, the payoff matrices $\mathbf{B}^{A}, \mathbf{L}^{A}\in \mathbb{R}^{2^n\times 2^n}$ can be decomposed as,
\begin{equation}
\mathbf{B}^{A}=\mathbf{Q}\mathbf{D}^B\mathbf{Q}^T,\text{ }\mathbf{L}^{A}=\mathbf{Q}\mathbf{D}^L\mathbf{Q}^T\mathbf{P}, 
\end{equation} 
where $\mathbf{D}^B, \mathbf{D}^L\in \mathbb{R}^{2^n\times 2^n}$ are the diagonal matrices with 
\begin{equation*}
 \mathbf{D}_{\sigma(A),\sigma(A)}^B=B^c_a(A),\text{ }\mathbf{D}_{\sigma(A),\sigma(A)}^L=L^c_a(A), \forall A\in 2^{[n]}.
\end{equation*}
The $\mathbf{Q}, \mathbf{P}\in \mathbb{R}^{2^n\times 2^n}$ are binary matrices: $\forall A,D\in 2^{[n]}$,
\begin{equation*}
\mathbf{Q}_{\sigma(A),\mu(D)}=\mathbbm{1}\{D^c \subseteq A\},\text{ }\mathbf{P}_{\sigma(A),\mu(D)}=\mathbbm{1}\{A=D\}. 
\end{equation*}
The notation $\mathbbm{1}\{\cdot\}$ is the indictor function.
\end{lemma}

The following lemma constructs the relation between the  matrices $\mathbf{B}^{A}, \mathbf{L}^{A}$ and the corresponding sub-matrices: $\mathbf{B}^{a}, \mathbf{L}^{a}$. 
\begin{lemma}\label{lm:completeandpartial}
\emph{(Complete matrix and sub-matrix)} The payoff matrices  $\mathbf{B}^{a}, \mathbf{L}^{a}\in \mathbb{R}^{N_a\times N_d}$ can be expressed as
\begin{equation}
\mathbf{B}^{a}=\mathbf{S}\mathbf{B}^{A}\mathbf{R},\text{ }, \mathbf{L}^{a}=\mathbf{S}\mathbf{L}^{A}\mathbf{R},
\end{equation}
where matrix $\mathbf{S}\in\mathbb{R}^{N_a\times 2^n}$ and $\mathbf{R}\in\mathbb{R}^{2^n\times N_d}$ are the block matrices, defined as,
\begin{align*}
&\mathbf{S}_{\sigma(A),\mu(U)}=\mathbbm{1}\{A=U^c\}, \forall A\in\mathcal{A}, U\in 2^{[n]},\\
&\mathbf{R}_{\sigma(U),\mu(D)}=\mathbbm{1}\{D=U^c\}, \forall D\in\mathcal{D}, U\in 2^{[n]}.
\end{align*}
\end{lemma}

Intuitively, the matrices $\mathbf{S}$ and $\mathbf{R}$ in Lemma~\ref{lm:completeandpartial} plays the role of extracting the rows and columns of the matrices $\mathbf{B}^{A}$ and $ \mathbf{L}^{A}$, whose indices belong to the feasible pure strategies of attacker and defender. For example, suppose that the defender's pure strategy space is a uniform matroid, i.e., $\mathcal{D}=\{D\subseteq [n]||D|\leq k\}$, if the index function satisfies $\sigma(U_1)\leq\sigma(U_2), |U_1|\geq |U_2|$, which means that the index of attacker's (defender's) pure strategy is increasing (decreasing) with the decreasing of the cardinality of each strategy, then the payoff matrices $\mathbf{B}^{a}$ and $ \mathbf{L}^{a}$ comes from the bottom left of the matrices $\mathbf{B}^{A}$ and $ \mathbf{L}^{A}$. The block matrices $\mathbf{S}$ and $\mathbf{R}$ can be represented as 
\begin{equation*}
  \mathbf{S}=\begin{bmatrix}
  \mathbf{0}_{N_a\times (2^n-N_a)} & \mathbf{I}_{N_a\times N_a}
  \end{bmatrix}\text{,   }\mathbf{R}=\begin{bmatrix}
 \mathbf{I}_{N_d\times N_d}\\
  \mathbf{0}_{(2^n-N_d)\times N_d}
  \end{bmatrix},
\end{equation*}
where $\mathbf{I}$ is the identity matrix.

Combining the results of Lemma~\ref{lm:decomplete} and Lemma~\ref{lm:completeandpartial}, we have the following decomposition of the payoff matrix $\mathbf{M}^a$.
\begin{theorem}\label{thm:departial}
\emph{(Decomposition of the payoff matrix)}  The payoff matrix $\mathbf{M}^a=\mathbf{B}^a+\mathbf{L}^a$ can be decomposed as 
\begin{equation}
	\mathbf{M}^a=\mathbf{E}(\mathbf{D}^b\mathbf{J}+\mathbf{D}^l\mathbf{K}),
\end{equation}	
where $\mathbf{D}^b, \mathbf{D}^l\in \mathbb{R}^{N_a\times N_a}$ are the diagonal matrices with 
\begin{equation*}
 \mathbf{D}_{\sigma(A),\sigma(A)}^b=B^c_a(A),\text{ }\mathbf{D}_{\sigma(A),\sigma(A)}^l=L^c_a(A), \forall A\in \mathcal{A}.
\end{equation*}
The $\mathbf{E}\in \mathbb{R}^{N_a\times N_a}$ and $\mathbf{J},\mathbf{K}\in \mathbb{R}^{N_a\times N_d}$are binary matrices:
\begin{align}
   &\mathbf{E}_{\sigma(A),\sigma(U)}=\mathbbm{1}\{U\subseteq A\}, \forall A,U\in \mathcal{A}; \mathbf{J}_{\sigma(A),\mu(D)}=\mathbbm{1}\{A\subseteq D^c\},\notag\\
   &\mathbf{K}_{\sigma(A),\mu(D)} =\mathbbm{1}\{A\subseteq D\}, \forall A\in \mathcal{A}, D\in \mathcal{D}.\notag
 \end{align}
\end{theorem}

As can be seen in Theorem~\ref{thm:departial}, we decompose the original exponential large payoff matrix $\mathbf{M}^a$ into the summation and the product of several simple matrices including binary matrices $\mathbf{E},\mathbf{J}, \mathbf{K}$ and two polynomial-sized diagonal matrices $\mathbf{D}^b$ and $\mathbf{D}^l$. Moreover, such a decomposition has a closed-form expression and the elements in those simple matrices can be implicitly represented. 

Based on the above decomposition results, we can let elementary matrices $\mathbf{E}_1=\mathbf{E}_2=\mathbf{E}$,  $\mathbf{F}_1=\mathbf{J}$ and $\mathbf{F}_2=\mathbf{K}$, and the corresponding affine transformation $f(\mathbf{p})=\mathbf{E}^T\mathbf{p}$ and $g_1(\mathbf{q})=\mathbf{J}\mathbf{q}$, $g_2(\mathbf{q})=\mathbf{K}\mathbf{q}$ to yield two polytopes: $\Delta^{a}_{N_a}=\{f(\mathbf{p})|\mathbf{p}\in\Delta_{N_a}\}$ and $\Delta^{d}_{N_d}=\{(g_1(\mathbf{q}),g_2(\mathbf{q}))|\mathbf{q}\in\Delta_{N_d}\}$. Then we can represent the minimax problem (\ref{eq:orgmodel}) as
\begin{equation}\label{eq:cptmodel}
  \min\limits_{(\bar{\mathbf{q}}_1,\bar{\mathbf{q}}_2)\in \Delta_{N_d}^d}\max\limits_{\bar{\mathbf{p}}\in \Delta_{N_a}^a} \bar{\mathbf{p}}^T(\mathbf{D}^b\bar{\mathbf{q}}_1+\mathbf{D}^l\bar{\mathbf{q}}_2),
\end{equation}

Based on the definition of our support set $S$ and matrices $\mathbf{D}^b$, $\mathbf{D}^l$, only the variables with indices belonging to $\sigma(S)$ has non-zero coefficients. Therefore, we can eliminate those variables with zero coefficients in (\ref{eq:cptmodel}) and project the polytopes $\Delta_{N_a}^a$ and $\Delta_{N_d}^d$ into the coordinates with indices belonging to $\sigma(S)$. The further simplified model can be expressed as
\begin{align}
 \textbf{Com} &\textbf{pact Minimax Problem}\notag\\
 &\min\limits_{(\bar{\mathbf{q}}_1,\bar{\mathbf{q}}_2)\in H_d}\max\limits_{\bar{\mathbf{p}}\in H_a} \bar{\mathbf{p}}^T(\widetilde{\mathbf{D}}^b\bar{\mathbf{q}}_1+\widetilde{\mathbf{D}}^l\bar{\mathbf{q}}_2),\label{eq:cptmodel1}
\end{align}
where\footnote{Note that each vector in $\Delta_{N_d}^d$ is consists of two parts $g_1(\mathbf{q})$ and $g_2(\mathbf{q})$. Here the correponding low-dimensional point is $(\pi_S(g_1(\mathbf{q}),$ $\pi_S(g_2(\mathbf{q}))$.} $H_a=\Pi_{S}(\Delta_{N_a}^a)$, $H_d=\Pi_{S}(\Delta_{N_d}^d)$, matrix $\widetilde{\mathbf{D}}^b$ and $\widetilde{\mathbf{D}}^l$ is obtained by extracting the non-zero columns and rows of matrix $\mathbf{D}^b$ and $\mathbf{D}^l$.

Since the size of the support set $|S|\leq N_a$ and $N_a=\text{poly}(n)$, we arrive at a compact representation of zero-sum security game with only poly$(n)$ variables. Note that in the above compact representation framework, the affine transformation $f_1$ and $f_2$ is the same as in our compact representation. The following theorem guarantees the correctness of our compact representation.
\begin{theorem}\label{thm:cpt}
\emph{(Correctness of compact representation)} $(\mathbf{p}^*,\mathbf{q}^*)$ is a Nash equilibrium of zero-sum security game  if and only if $(\pi_S(f(\mathbf{p}^*)),$ $(\pi_S(g_1(\mathbf{q}^*)),\pi_S(g_2(\mathbf{q}^*)))$ is the optimal solution of compact minimax problem  (\ref{eq:cptmodel1}).
\end{theorem}

\begin{remark}
Based the specific form of (\ref{eq:cptmodel1}), the complexity of obtaining the above compact representation (\ref{eq:cptmodel1}) is dependent on the complexity of obtaining matrices $\widetilde{\mathbf{D}}^b$, $\widetilde{\mathbf{D}}^l$ and the  projected polytopes $H_a$, $H_d$. The non-zero elements of matrices $\widetilde{\mathbf{D}}^b$ and $\widetilde{\mathbf{D}}^l$ are the common utilities, which can be calculated in time $O(n^c)$ by Definition~\ref{def:com}. The complexity of representing $H_a$ and $H_d$ is dependent on their corresponding description length.
\end{remark}

\subsection{Linear Programming Approach}

The compact minimax problem has a linear objective function. If the polytope $H_a$ and $H_d$ is convex, such a problem can be cast into a linear programming approach. The following technical lemma shows that the above transformed and projected polytope is convex.
\begin{lemma}\label{lm:convex}
The polytopes $H_a$ and $H_d$ is convex.
\end{lemma}
Based on Lemma~\ref{lm:convex}, we can formulate the minimax problem by following equivalent linear programming model,
\begin{align}
\textbf{Comp} & \textbf{act Linear Programming}\notag\\
&\min\quad u\label{eq:compact}\\
& \begin{array}{r@{\quad}l@{}l@{\quad}l}
s.t. &\mathbf{v}^T(\widetilde{\mathbf{D}}^b\bar{\mathbf{q}}_1+\widetilde{\mathbf{D}}^l\bar{\mathbf{q}}_2)\leq u, \forall\mathbf{v}\in I_a,\qquad \notag\\
 &(\bar{\mathbf{q}}_1,\bar{\mathbf{q}}_2)\in H_d,
\end{array} 
\end{align}
where $I_a$ denotes the set of vertices of the convex polytope $H_a$.

In the sequel, we refer to the \textbf{compact problem} as the above linear programming problem. Based on our compact representation, a natural question that arises is whether can we efficiently solve such a linear programming problem and implement the optimal solution by the defender's mixed strategy? We will answer this question in the next section. 

\section{Solving the Zero-sum Security Game is a Combinatorial Problem}

In this section, we will build the connection between the equilibrium computation in the zero-sum security game and the following defender oracle problem (DOP) via our compact representation framework. For simplicity, we use $I_d$ to denote set of vertices of the convex polytope $H_d$.
\begin{definition}
 \emph{(Defender oracle problem)} For any given vector $\mathbf{w}\in\mathbb{R}^{2|S|}$, determine,
\begin{equation}\label{eq:dopzerosum}
\mathbf{x}^*=\arg\min_{\mathbf{x}\in I_d} \mathbf{w}^T\mathbf{x}.  
\end{equation}
\end{definition}

The main result of this section is the following theorem.
\begin{theorem}\label{thm:organddop}
\emph{(NE computation and denfender oracle problem)} There is a \emph{poly}$(n)$ time algorithm to compute the defender's Nash equilibrium (strong Stackelberg equilibrium), \textbf{if and only if} there is a \emph{poly}$(n)$ time algorithm to compute the defender oracle problem.
\end{theorem}

To obtain above reduction, we adopt the following path: we first show how the compact problem and the defender oracle problem can be reduced to each other in poly($n$) time; then we exploit the geometric structure of polytope $H_a$ and $H_d$ to construct two poly($n$) time vertex mapping algorithms to obtain the reduction between the equilibrium computation and the compact problem.

\subsection{Reduction between Compact Problem and DOP}

The linear programing problem (\ref{eq:compact}) has $\text{poly}(n)$ number of variables and possibly exponentially many constraints due to the irregularity of the polytope $H_d$. Therefore, we can apply the ellipsoid method to solve such an LP, given a $\text{poly}(n)$ time separation oracle. Specifically, the separation oracle of such LP is defined as following.
\begin{definition}
\emph{(Separation oracle for LP (\ref{eq:compact}))} For any given $(\bar{\mathbf{q}}_1,\bar{\mathbf{q}}_2)$ and $\bar{u}$, either $(\bar{\mathbf{q}}_1,\bar{\mathbf{q}}_2)$ and $\bar{u}$ satisfy all the constraints of (\ref{eq:compact}) or finds a hyperplane:
\begin{align*}
  &\mathbf{a}_1^T\bar{\mathbf{q}}_1+\mathbf{a}_2^T\bar{\mathbf{q}}_2+b\bar{u}> \mathbf{a}_1^T\mathbf{q}_1+\mathbf{a}_2^T\mathbf{q}_2+bu,
\end{align*}
$\forall (\mathbf{q}_1, \mathbf{q}_2), u \text{ satisfy constraints of }(\ref{eq:compact})$
\end{definition}

Similarly, the separation oracle of the defender oracle problem is defined as following.
\begin{definition}
\emph{(Separation oracle for DOP (\ref{eq:dopzerosum}))} For any given $(\bar{\mathbf{q}}_1,\bar{\mathbf{q}}_2)$, either $(\bar{\mathbf{q}}_1,\bar{\mathbf{q}}_2)$ belongs to $H_d$, or finds a hyperplane: $\forall (\mathbf{q}_1,$ $\mathbf{q}_2)\in H_d$ such that
\begin{align*}
  &\mathbf{a}_1^T\bar{\mathbf{q}}_1+\mathbf{a}_2^T\bar{\mathbf{q}}_2> \mathbf{a}_1^T\mathbf{q}_1+\mathbf{a}_2^T\mathbf{q}_2 .
\end{align*}
\end{definition}

Based on Theorem~\ref{thm:equioptandsep}, both our compact problem (\ref{eq:compact}) and defender oracle problem is equivalent to their corresponding separation oracle problem. To obtain the equivalence between the compact problem and the defender oracle problem, it remains to show the equivalence between the above two separation oracles. 

We first examine one direction: the separation oracle of such an LP can be reduced to the separation oracle of DOP. Specifically, the separation oracle of such an LP can be reduced to the following two parts: given any $(\bar{\mathbf{q}}_1,\bar{\mathbf{q}}_2)$ and $\bar{u}$, (1) \emph{membership problem}: decide whether $(\bar{\mathbf{q}}_1,\bar{\mathbf{q}}_2)$ $ \in H^d$. If not, generate a hyperplane that separating $(\bar{\mathbf{q}}_1,\bar{\mathbf{q}}_2)$, $\bar{u}$ from $H^d$; (2) \emph{inequality constraint problem}: decide whether all the inequality constraints hold. If not, find one violating constraint. We have the following result for these problems. 
\begin{lemma}\label{lm:miop}
	The \emph{membership problem} and the \emph{inequality constraint problem} of  LP (\ref{eq:compact}) can be reduced to the separation oracle of the defender oracle problem in \emph{poly}($n$) time.
\end{lemma}

The basic idea in the proof of Lemma~\ref{lm:miop} is to show that the number of vertices of $H_a$ is poly($n$) and each vertex has a close-form expression. Then we can implicitly check the \emph{inequality constraint problem} in poly($n$) time, and the \emph{membership problem} is indeed the separation problem for the DOP. The reverse direction is guaranteed by the following lemma.

\begin{lemma}\label{lm:zerosumdopoandcpto}
	The separation oracle of the defender oracle problem can be reduced to the separation oracle of  LP (\ref{eq:compact}) in \emph{poly}($n$) time.
\end{lemma}
The idea of the proof of Lemma~\ref{lm:zerosumdopoandcpto} is that, for any input $(\bar{\mathbf{q}}_1,\bar{\mathbf{q}}_2)$ of separation oracle for DOP, we can choose a specific $u_0\in\mathbb{R}$ such that
\begin{equation*}
  u_0=|S|(\|\widetilde{\mathbf{D}}^b\|+\|\widetilde{\mathbf{D}}^l\|) 
\end{equation*}
where $\|\cdot\|$ is the matrix's spectrum norm. Then we take $(\bar{\mathbf{q}}_1,\bar{\mathbf{q}}_2)$ and $u_0$ as the input of the separation oracle for the compact problem. If the output is yes, we also get a certificate that $(\bar{\mathbf{q}}_1,\bar{\mathbf{q}}_2)\in H_d$; if not, output a $(2|S|+1)-$dimensional separating hyperplane: 
\begin{align*}
  &\mathbf{a}_1^T\bar{\mathbf{q}}_1+\mathbf{a}_2^T\bar{\mathbf{q}}_2+bu_0> \mathbf{a}_1^T\mathbf{q}_1+\mathbf{a}_2^T\mathbf{q}_2+bu,
\end{align*}
$\forall (\mathbf{q}_1, \mathbf{q}_2), u$ $ \text{satisfy constraints of } (\ref{eq:compact})$. Due to the specific choice of $u_0$, we can show that such $\mathbf{a}_1^T\bar{\mathbf{q}}_1+\mathbf{a}_2^T\bar{\mathbf{q}}_2$ also forms a separating hyperplane for all $(\mathbf{q}_1, \mathbf{q}_2) \in H_d$. The intuition behind this step is that when we choose a large enough $u_0$, the feasible region defined by the constraints in LP ($\ref{eq:compact}$) will degenerate to the polytope $H_d$.

Based on Lemma~\ref{lm:miop} and Lemma~\ref{lm:zerosumdopoandcpto}, we arrive at the reduction between the compact problem and the defender oracle problem.

\subsection{Reduction between Equilibrium  Computation and Compact Problem}

To obtain the reduction between the equilibrium computation and the compact problem, there exist two issues: first, how to transform the input instance of each problem to the other one in poly($n$) time; second, how to map the optimal solution of each problem to the other in poly($n$) time. Since the input of the equilibrium computation problem are the uitlity functions $\{B_a(U)\}$ and $\{L_a(U)\}$ and the input of compact problem are the common utilities $\{B_a^c(U)\}$ and $\{L_a^c(U)\}$ (all the elements of matrices $\mathbf{D}^b$ and $\mathbf{D}^l$ are the common utilities), such transformation can be completed in $O(2^cn^c)=$ poly($n$) time based on Definition~\ref{def:com} and Lemma~\ref{lm:com}.

To resolve the second issue, we first consider how to map the optimal solution of compact problem to the defender's optimal mixed strategies. Based on Theorem~\ref{thm:convdecomp}, we obtain that if the separation problem of LP (\ref{eq:compact}) can be solved in $\text{poly}(n)$ time, we can decompose any feasible point $\mathbf{x}$ into a convex combination of at most $(2|S|+1)$ vertices of the polytope defined by those constraints. Note that this is precisely the DOP required for above reduction. Applying this result to the optimal solution $(\mathbf{q}_1^*,\mathbf{q}_2^*)$ of the LP (\ref{eq:compact}), we can get a convex decomposition that 
\begin{equation}
(\mathbf{q}_1^*,\mathbf{q}_2^*)=\sum_{i=1}^{2|S|+1}\lambda_i (\mathbf{v}^i_1,\mathbf{v}^i_2),  
\end{equation}
where $(\mathbf{v}^i_1,\mathbf{v}^i_2)\in I_d$. The basic fact is that the defender's mixed strategy can be regarded as a convex combination of its pure strategies, each of which corresponds to a vertex of simplex $\Delta_{N_d}$. If we can map the vertices $(\mathbf{v}^i_1,\mathbf{v}^i_2)$ back to the vertices (pure strategy) of the original game, denoted by $h((\mathbf{v}^i_1,\mathbf{v}^i_2))$, the mixed strategies of the defender can be expressed as 
\begin{equation}
	\mathbf{q}^*=\sum_{i=1}^{2|S|+1}\lambda_i h((\mathbf{v}^i_1,\mathbf{v}^i_2)).
\end{equation}
Thus, the key lies in how to compute $h((\mathbf{v}^i_1,\mathbf{v}^i_2))$ in $\text{poly}(n)$ time. 

To tackle this problem, we need to investigate the geometric structure of polytope $H_d$. First, considering an arbitrary defender's pure strategy $D\in \mathcal{D}$, the corresponding vertex in $\Delta_{N_d}$ is a unit vector $\mathbf{e}^D\in \mathbb{R}^{N_d}$ with only one non-zero element $\mathbf{e}^D_{\mu(D)}=1$. Based on the definition of the transformation $g_1(\mathbf{q})$ and $g_2(\mathbf{q})$, the corresponding point of polytope $H_d$ is 
\begin{equation}
	(g_1(\mathbf{e}^D),g_2(\mathbf{e}^D))=(\mathbf{J}\mathbf{e}^D,\mathbf{K}\mathbf{e}^D)=(\mathbf{J}_{\mu(D)},\mathbf{K}_{\mu(D)}),
\end{equation}
where $\mathbf{J}_{\mu(D)}$ and $\mathbf{K}_{\mu(D)}$is the $\mu(D)$th column of matrix $\mathbf{J}$ and $\mathbf{K}$. Then the corresponding point $\mathbf{v}^D$ of the projected polytope $H^d$ is 
\begin{equation}
	\mathbf{v}^D=\left(\pi_{S}(\mathbf{J}_{\mu(D)}),\pi_{S}(\mathbf{K}_{\mu(D)})\right),
\end{equation}
which is the sub-vector of $\mathbf{J}_{\mu(D)}$ and $\mathbf{K}_{\mu(D)}$. The problem is that the vertex in the high-dimensional polytope may not project to a vertex of its low-dimensional image. However, the following lemma will provide a positive result.
\begin{lemma}\label{lm:cptvertex}
\emph{(Geometric structure of $H_d$)} For any support set $[n]\in S\in\mathcal{A}$, the vertices of the polytope $H_d$ are the columns of the sub-matrix of $\begin{bmatrix}
\mathbf{J}\\
\mathbf{K}    
  \end{bmatrix}$, which is formed by extracting the row whose index belongs to $\sigma(S)$.
\end{lemma}

Since we have a closed-form expression of the matrix $\mathbf{J}$ and $\mathbf{K}$, we can construct a vertex mapping algorithm from low-dimensional vertex to the defender's pure strategy. The efficiency and the correctness of Algorithm~\ref{alg1} is justified by following lemma.

\begin{algorithm}[t]
\renewcommand{\algorithmicrequire}{ \textbf{Input:}} 
\renewcommand{\algorithmicensure}{ \textbf{Output:}} 
\caption{Vertex Mapping from Vertex to Pure Strategy}   
\label{alg1}  
\begin{algorithmic}  
\REQUIRE Vertex $(\mathbf{v}_1,\mathbf{v}_2)\in I^d$  
\ENSURE Defender's pure strategy $D$.
\STATE $T=\emptyset$;
\FOR{each $i \in [n]$} 
\STATE Examine each coordinate of vertex: \\ \textbf{if} $\mathbf{v}_{1,\sigma(\{i\})}\neq 0$ \textbf{ then} $T=T\cup\{i\}$;
\ENDFOR
\STATE $D=T^c$;
\end{algorithmic}  
\end{algorithm}

\begin{lemma}\label{lm:cpt2org}
\emph{(Correctness of vertex mapping algorithm)}	The vertex mapping algorithm~\ref{alg1} runs in $O(n)$ time and maps each vertex of $H_d$ to a unique pure strategy.
\end{lemma}
Note that our vertex mapping algorithm only examines $n$ instead of all the coordinates of each vertex of $H_d$ to recover a defender's pure strategy. The reason behind this result is that there exists a one-one correspondence between each pure strategy and those $n$ coordinates of each vertex of polytope $H_d$. Intuitively, those $n$ coordinates of each vertex of $H_d$ is binary and therefore there exists possibly $2^n$ possibilities, each of which corresponds to a pure strategy. 

The other direction follows from the following argument. Suppose that the problem of equilibrium computation is solved in poly ($n$) time and the optimal defender's mixed strategy is denoted by $\mathbf{q}^*$. Invoking a known result in game theory (Theorem 4 in~\cite{lipton2003playing}), the number of non-zero probability of the Nash equilibrium is less than the rank of the payoff matrix. Since the rank of payoff matrix $\mathbf{M}^a$ is $O(n^c)$, the number of non-zero coordinates in $\mathbf{q}^*$ is at most $O(n^c)=$ poly($n$) and $\mathbf{q}^*$ can be expressed as
\begin{equation}
  \mathbf{q}^*=\sum\limits_{i=1}^{\text{poly}(n)}\lambda_i \mathbf{e}^i.
\end{equation}

Therefore, we can determine the optimal solution of the compact problem in poly$(n)$ time by constructing the following poly$(n)$ time vertex mapping algorithm from a pure strategy $\mathbf{e}^i$ to a vertex of $H_d$.
\begin{algorithm}  
\renewcommand{\algorithmicrequire}{ \textbf{Input:}} 
\renewcommand{\algorithmicensure}{ \textbf{Output:}} 
\caption{Vertex Mapping from Pure Strategy to Vertex}     
\label{alg:pure2vertex}
\begin{algorithmic}  
\REQUIRE Defender's Pure Strategy $D$ 
\ENSURE Vertex $\mathbf{v}^D\in I_d$
\FOR{each $V \in \mathcal{A}$} 
\STATE \textbf{if} $V\subseteq D^c$ \textbf{ then} $\mathbf{v}^D_{1,\sigma(V)}=1$; \textbf{else } $\mathbf{v}^D_{1,\sigma(V)}=0$.
\STATE \textbf{if} $V\subseteq D$ \textbf{ then} $\mathbf{v}^D_{2,\sigma(V)}=1$; \textbf{else } $\mathbf{v}^D_{2,\sigma(V)}=0$.
\ENDFOR
\STATE Output vertex $\mathbf{v}^D=(\mathbf{v}^D_{1},\mathbf{v}^D_{2})$.
\end{algorithmic}  
\end{algorithm}

The intuition behind this result is similar to the previous vertex mapping algorithm and the correctness of Algorithm~\ref{alg:pure2vertex} is guaranteed by the following lemma.
\begin{lemma}\label{lm:org2cpt}
\emph{(Correctness of vertex mapping algorithm)}	Vertex mapping algorithm~\ref{alg:pure2vertex} runs in $O(n^c)$ time and maps each defender's pure strategy $D$ to a unique vertex of $H_d$.
\end{lemma}

Based on Lemma~\ref{lm:miop}, Lemma~\ref{lm:zerosumdopoandcpto}, Lemma~\ref{lm:cpt2org} and Lemma~\ref{lm:org2cpt}, we arrived the desired result in Theorem~\ref{thm:organddop}.

\section{Non-zero-sum Security Game}

In this section, we assume that the security game is non-zero-sum, in which the benefit $B_a(U)$ ($B_d(U)$)  of attacker (defender) may not equal to the the loss $L_d(U)$ ($L_a(U)$) of the defender (attacker). We consider the computation of two mostly adopted concepts including the strong Stackelberg equilibrium (SSE) and the Nash equilibrium (NE). For the SSE, we prove analogous equivalence theorem as the zero-sum case. For the NE, we relax the assumption that the attacker's resource limit $c$ is constant to an arbitrary number, but assuming the additive utility function. Then we prove our equivalence theorem for the problem of equilibrium computation.

\subsection{Strong Stackelberg Equilibrium}

The main result of this subsection is given by the following theorem.
\begin{theorem}\label{thm:sseanddop}
\emph{(SSE and DOP)} There is a \emph{poly}$(n)$ time algorithm to compute the defender's strong Stackelberg equilibrium, \textbf{if and only if} there is a \emph{poly}$(n)$ time algorithm to compute the defender oracle problem (\ref{eq:dopzerosum}).
\end{theorem}

The ``only if'' direction follows straightforwardly from Theorem~\ref{thm:organddop}. Considering the fact that the set of Nash equilibrium is equivalent to the set of strong Stackelberg equilibrium in the zero-sum game, and the zero-sum game is a special case of the non-zero-sum game, such a direction can be obtained by converse. For the ``if'' direction, we need the following existing technical lemma, which shows that the SSE of the game can be computed by multiple linear programming approach.
\begin{lemma}
 (Multiple linear programming of SSE~\cite{conitzer2006computing}) Solving the strong Stackelberg equilibrium can be formulated as a multiple linear programming problem.
\begin{align}
 &\textbf{\emph{Multiple LP}}\notag\\
&\max\quad (\mathbf{e}^{A})^T\mathbf{M}^d\mathbf{q}\label{eq:multiLP}\\
& \begin{array}{r@{\quad}l@{}l@{\quad}l}
s.t. &(\mathbf{e}^{A})^T\mathbf{M}^a\mathbf{q}\geq(\mathbf{e}^{A'})^T\mathbf{M}^a\mathbf{q} , \forall A'\in \mathcal{A},\qquad \notag\\
 &\mathbf{q}\in \Delta_{N_d},
\end{array} 
\end{align}
Solving above linear programming for all $A\in\mathcal{A}$, then picking the optimal defender's mixed strategy of the LP with the largest objective value.
\end{lemma}

The intuition behind this result is that, in linear programming problem (for each $A\in\mathcal{A}$), the defender optimizes his mixed strategy under the constraint that the attacker's best response is $A$. Once we have solved these LPs for all the attacker's best response, we compare all the optimal mixed strategies and choose the best one, which is also the optimal solution overall (without constraints on which is the best response). 

Since the attacker's resource limit $c$ is constant, there exists poly ($n$) number of LPs in the above approach, but there still exists exponential number of variables in each LP. Fortunately, we can still follow our transformation and projection approach in the previous zero-sum scenario. Specifically, based on the same argument of Theorem~\ref{thm:departial}, we can decompose both attacker's and defender's payoff matrices as
\begin{equation}
 \mathbf{M}^a=\mathbf{E}(\mathbf{D}_{a}^b\mathbf{J}+\mathbf{D}_{a}^l\mathbf{K})\quad\text{and}\quad\mathbf{M}^d=\mathbf{E}(\mathbf{D}_{d}^b\mathbf{J}+\mathbf{D}_{d}^l\mathbf{K}),
\end{equation}
and choose the same transformation such that $f(\mathbf{p})=\mathbf{E}^T\mathbf{p}$ and $g_1(\mathbf{q})=\mathbf{J}\mathbf{q}$, $g_2(\mathbf{q})=\mathbf{K}\mathbf{q}$. Then we eliminate the variables with zero-coefficients and project the polytope into corresponding low-dimensional image. Therefore, we can formulate above multiple linear programming problem as following compact problem with only poly($n$) variables.
\begin{align}
&\textbf{Compact Multiple LP}\notag\\
&\max\text{  } \mathbf{v}^T(\widetilde{\mathbf{D}}^b_d\bar{\mathbf{q}}_1+\widetilde{\mathbf{D}}^l_d\bar{\mathbf{q}}_2)\label{eq:cptmultiLP}\\
& \begin{array}{r@{\quad}l@{}l@{\quad}l}
s.t. &\mathbf{v}^T(\widetilde{\mathbf{D}}^b_a\bar{\mathbf{q}}_1+\widetilde{\mathbf{D}}^l_d\bar{\mathbf{q}}_2)\geq \bar{\mathbf{v}}^T(\widetilde{\mathbf{D}}^b_a\bar{\mathbf{q}}_1+\widetilde{\mathbf{D}}^l_d\bar{\mathbf{q}}_2) , \forall \bar{\mathbf{v}}\in I_a,\qquad \notag\\
 &(\bar{\mathbf{q}}_1,\bar{\mathbf{q}}_2)\in H_d,
\end{array} 
\end{align}
where the matrix $\widetilde{\mathbf{D}}^b_d$ and the other diagonal matrices are obtained by extracting the non-zero columns and the rows of matrix $\mathbf{D}^b_d$. Clearly, such a compact problem can be reduced to the defender oracle problem (\ref{eq:dopzerosum}) and the optimal solution of compact problem can be mapped to a defender's optimal mixed strategy based on a similar argument with Lemma~\ref{lm:cptvertex} and Lemma~\ref{lm:cpt2org}. Thus, we arrive our desired results in Theorem~\ref{thm:sseanddop}.

\subsection{Nash Equilibrium}

It is well known that computing the Nash equilibrium of a two-player normal form game is PPAD-hard~\cite{chen2009settling,daskalakis2009complexity}. In the security game, we can potentially compute the Nash equilibrium in poly($n$) time in the case of the single attacker resource~\cite{xu2016mysteries}, because we can transform the non-zero-sum game into an equivalent zero-sum game~\cite{korzhyk2011stackelberg}.  However, when the attacker has multiple resources, the solvability is still an open problem. In this subsection, we assume that the attacker's resource limit $c$ is an arbitrary number instead of a constant, and all the utility functions are additive. The main result of this subsection is given by the following theorem.

\begin{theorem}\label{thm:dopandnonzero}
 \emph{(Non-zero-sum NE computation and DOP)} There is a \emph{poly}($n$) time algorithm to compute the defender's Nash equilibrium, \textbf{if and only if} there is a \emph{poly}($n$) time algorithm to compute the defender oracle problem: for any given $\mathbf{w}\in\mathbb{R}^n$
\begin{equation}
  \arg\min\limits_{\mathbf{x}\in I_d} \mathbf{w}^T\mathbf{x}.
\end{equation}
\end{theorem} 

The ``only if '' direction can be obtained by the converse and Theorem~\ref{thm:organddop}, thus we only focus on the how to compute the defender's Nash equilibrium in poly$(n)$ time given a poly$(n)$ time defender oracle. In this case, there exists two key challenges: first, not only is the defender's pure strategy space exponential large, but also the attacker's pure strategy space; second, the Nash equilibrium in non-zero-sum game may not be the corresponding minimax equilibrium when the attacker has multiple resources~\cite{korzhyk2011stackelberg}, thus we cannot apply the linear programming approach to straightforwardly solve such a problem.

The following lemma exhibits a crucial property of common utility when all the utility function is additive such that all the common utility is equal to zero except those defined on the singleton set.
\begin{lemma}\label{lm:add}
\emph{(Additive utility function)} If the attacker's utility functions are additive such that $B_a(U)=\sum_{i\in U}B_a(\{i\})$ and $L_a(U)=\sum_{i\in U}L_a(\{i\})$ for all $U\in 2^{[n]}$,  then the corresponding common utilities satisfy
\begin{equation}
B_a^c(U)=L_a^c(U)=0, \text{ if } |U|>1.
\end{equation}
\end{lemma}
A similar result holds for defender's utility functions $B_d(\cdot)$, $L_d(\cdot)$ and their common utilities $B_d^c(\cdot)$, $L_d^c(\cdot)$. Note that the common utility function is indeed the original utility function when $U$ is a singleton set. Then, according to the definition of matrix $\mathbf{D}_{a}^b$ and $\mathbf{D}_{a}^l$, we can observe that although they are exponential large matrices, there exists only $n$ non-zero elements in the main diagonal. Therefore, the support set $S=[n]$ based on the definition, and we can define the following $n-$dimensional polytope.
\begin{equation}
 H_a'=\Pi_{[n]}(\Delta_{N_a}^a), \text{ } H_d'=\{\pi_{[n]}(\mathbf{J}\mathbf{q}), \forall \mathbf{q}\in \Delta_{N_d}\}.
\end{equation}
which projects our transformed polytopes to the coordinates whose indices belong to the singleton set. Note that, in this case, we only use the vector $\mathbf{J}\mathbf{q}$ instead both of $\mathbf{J}\mathbf{q}$ and $\mathbf{K}\mathbf{q}$ to form the transformed and projected polytope. The reason is that there exists a linear coupling between $\pi_{[n]}(\mathbf{J}\mathbf{q})$ and $\pi_{[n]}(\mathbf{K}\mathbf{q})$. The formal description can be seen in Lemma~\ref{lm:ncptvertex}. The following result provides a compact representation of the non-zero-sum security game.
\begin{lemma}\label{lm:nonzerosdd}
\emph{(Compact representation of non-zero-sum security game)} The strategy profile ($\mathbf{p}^*,\mathbf{q}^*$)  is a Nash equilibrium of the non-zero-sum security game if and only if
\begin{equation}\label{eq:nonzerosdd}
  \begin{aligned}
  &U'_a(\mathbf{a}^*,\mathbf{t}^*)\geq U_a'(\mathbf{a},\mathbf{t}^*), \forall \mathbf{a}\in H_a', \\
  &U'_d(\mathbf{a}^*,\mathbf{t}^*)\geq U_d'(\mathbf{a}^*,\mathbf{t}), \forall \mathbf{t}\in H_d'.
\end{aligned}
\end{equation}
where\footnote{For simplicity, we write $\mathbf{a}_{\sigma(\{i\})}$ and $\mathbf{t}_{\mu(\{i\})}$ as $a_i$ and $t_i$.} 
\begin{align*}
 &U'_a(\mathbf{a},\mathbf{t})=\sum\limits_{i=1}^{n}a_i\left[t_{i} L_a(i)+(1-t_i)B_a(i)\right],\\
&U'_d(\mathbf{a},\mathbf{t})=\sum\limits_{i=1}^{n}a_i\left[t_{i} B_d(i)+(1-t_i)L_d(i)\right].
\end{align*} 
\end{lemma}
 
As can be seen in Lemma~\ref{lm:nonzerosdd}, the strategy profile $(\mathbf{p}^*,\mathbf{q}^*)$ is a Nash equilibrium in non-zero-sum security game if and only if its low dimensional image $(\mathbf{a}^*,\mathbf{t}^*)$ is also an equilibrium point. Based on the definition of matrix $\mathbf{E}$ and $\mathbf{J}$, we have
\begin{align*}
 &a_i=\sum\limits_{A\in\mathcal{A}:i\in A}\mathbf{p}_{\sigma (A)},\text{ } t_i=\sum\limits_{D\in\mathcal{D}:i\in D}\mathbf{q}_{\mu (D)},
\end{align*}
which can be regarded as the marginal probability that target $i$ is attacked and defended among the strategy profile $(\mathbf{p},\mathbf{q})$.

\begin{remark}
Lemma~\ref{lm:nonzerosdd} provides a compact representation of the non-zero-sum security game, and the objective function $U_a'$ and $U_d'$ are quite similar to the the one used in~\cite{kiekintveld2009computing}. However, the previous result is limited to the assumption that the defender's pure strategy space is a uniform matroid. However, our result is more general and does not dependent on the structure of $\mathcal{D}$.
\end{remark}

Although we have a polynomial-sized representation, it is still a non-zero-sum ``game'' and its optimal solution may not correspond to its minimax solution. Fortunately, inspired by the transformation introduced in~\cite{korzhyk2011stackelberg}, we can exploit the specific structure of the above problem to transform the compactly represented problem into a $n-$dimensional saddle point problem.

\begin{lemma}\label{lm:zerosumsdd}
\emph{(Saddle point problem)} The ($\mathbf{a}^*,\mathbf{t}^*$) satisfy condition (\ref{eq:nonzerosdd}) if the ($h(\mathbf{a}^*),$ $\mathbf{t}^*$) is a saddle point of following problem,
\begin{equation}\label{eq:zerosdd}
  \begin{aligned}
  &U'_a(h(\mathbf{a}^*),\mathbf{t}^*)\geq U_a'(h(\mathbf{a}),\mathbf{t}^*),\forall \mathbf{a}\in H_a'\\
  &U'_a(h(\mathbf{a}^*),\mathbf{t}^*)\leq U_a'(h(\mathbf{a}^*),\mathbf{t}),\forall \mathbf{t}\in H_d',
\end{aligned}
\end{equation}
where the one-to-one transform function $h: \mathbb{R}^n\rightarrow \mathbb{R}^n$is defined as follows:
\begin{equation}
 h_i(\mathbf{a})=\frac{B_d(\{i\})-L_d(\{i\})}{B_a(\{i\})-L_a(\{i\})}a_i,
\end{equation}
\end{lemma}

Based on Lemma~\ref{lm:zerosumsdd}, we can first determine a saddle point of function $U_a'(\mathbf{a},\mathbf{t})$. Due to the fact that function $U_a'(h(\mathbf{a}),\mathbf{t})$ is linear in $\mathbf{a}$ and $\mathbf{t}$ and the polytope $H_a'$ and $H_d'$ is convex (Lemma~\ref{lm:convex}), we can apply the Sion's minimax theorem and solve such problem by the minimax problem,
\begin{equation}
 \min\limits_{\mathbf{t}\in H_d'}\max\limits_{\mathbf{a}\in H_a'}U_a'(h(\mathbf{a}),\mathbf{t}),
\end{equation}  
which can be further tackled by the following linear programming approach.
\begin{align}
&\min\quad u\label{eq:compactzero}\\
& \begin{array}{r@{\quad}l@{}l@{\quad}l}
s.t. &\sum\limits_{i=1}^{n}h_i(\mathbf{v})\left[t_{i} L_a(i)+(1-t_i)B_a(i)\right]\leq u, \forall \mathbf{v}\in I_a',\qquad \notag\\
 &\mathbf{t}\in H_d'.
\end{array} 
\end{align}

Similarly, such an LP can be reduced to the membership problem and inequality constraint problem. The membership problem can be reduced to the defender oracle problem (\ref{eq:dopzerosum}), and the only difference is that $H_d'$ is a $n-$dimensional polytope in this case. To solve the inequality constraint problem, we first examine the geometric structure of polytope $H_a'$.
\begin{lemma}\label{lm:hypercube}
 \emph{(Geometric structure of $H_a'$)} The polytope $H_a'$ is the intersection of a $n-$dimensional cube and $n-$dimensional hyperplane,
 \begin{equation}
   H_a'=\{\mathbf{x}\in\mathbb{R}^n|\sum\limits_{i=1}^{n}x_i\leq c, x_i\geq 0, \forall 1\leq i\leq n\}.
 \end{equation}
\end{lemma}
Since any linear program achieves optimality at some vertex of its feasible region, based on Lemma~\ref{lm:hypercube}, the inequality constraint problem can be further reduced to the following polynomial-sized linear programming problem.
\begin{align}
&\max \sum\limits_{i=1}^{n}h_i(\mathbf{v})\left[t_{i} L_a(i)+(1-t_i)B_a(i)\right]\label{eq:cptzeroineq}\\
& \begin{array}{r@{\quad}l@{}l@{\quad}l}
s.t. &\sum\limits_{i=1}^{n}v_i\leq c, v_i\geq 0, \forall 1\leq i\leq n.\notag
\end{array} 
\end{align}
which can be solved in the poly$(n)$ time by the interior point method. If the optimal value of the above linear programming is less than $u$, then all the inequality constraints are satisfied; if the optimal value is larger than $u$, we output the point $\mathbf{v}$ attains this optimal value, which corresponds to a violating constraint.

Now the remaining is to determine how to map such a saddle point back to the optimal defender's mixed strategy. Since our transformation is linear and the inverse function is given by
\begin{equation}
  h_i^{-1}(\mathbf{a})=\frac{B_a(\{i\})-L_a(\{i\})}{B_d(\{i\})-L_d(\{i\})}a_i,
\end{equation}
we can first decompose such saddle point into the convex combination of the vertices given a poly$(n)$ time separation oracle; then use the inverse function to map each vertex back to the vertex of polytope $H_d'$. The mapping from the the vertex of polytope $H_d'$ to a defender's pure strategy follows the Algorithm~\ref{alg1}.

Combining the result of Lemma~\ref{lm:nonzerosdd} and Lemma~\ref{lm:zerosumsdd}, we arrived the desired result in Theorem~\ref{thm:dopandnonzero}.

\begin{remark}
  The reduction from the equilibrium computation to the defender oracle problem does not require the assumption that the attacker's pure strategy space $\mathcal{A}$ is a uniform matroid. Indeed, if $\mathcal{A}$ encodes a polynomial solvable problem such as uniform matroid, bipartite matching and $2-$weighted cover, we can obtain a same result in Theorem~\ref{thm:dopandnonzero}; otherwise, the polynomial solvability of the equilibrium computation will depends on both $\mathcal{A}$ and $\mathcal{D}$.
\end{remark}

\section{Consequences and Applications}

In this section, we further investigate the defender oracle problem and provides various interesting applications of our theoretical framework.

\subsection{What is the Defender Oracle Problem}

Through a series of reductions, we determine that the security game with non-additive utility functions and multiple attacker resources is essentially a defender oracle problem defined on a low- dimensional polytope $H_d$. However, the complicated form of polytope $H_d$ still prevents us from uncovering the effect of our assumptions. The promising observation though is that since oracle problem attains its maximum in some vertices of $H_d$, we only need to examine the geometric structure of its vertices. 

First, considering the fact that each vertex $\mathbf{v}$ of polytope $H_d$ consists of two parts such that $\mathbf{v}=(\mathbf{v}_1, \mathbf{v}_2)$, and $\mathbf{v}_1$, $\mathbf{v}_2$ comes from two different transformations and same projection of the one vertex of polytope $\Delta_{N_d}$. Therefore, there exists a linear coupling between these two parts $\mathbf{v}_1$ and $\mathbf{v}_2$. The following lemma justifies this intuition and show that the coordinates of $\mathbf{v}_1$ and $\mathbf{v}_2$ whose indices belong to those singleton sets exhibits a complementary relation.
\begin{lemma}\label{lm:ncptvertex}
\emph{(Geometric structure of $H_d$)} For any vertex $(\mathbf{v}_1,\mathbf{v}_2)$ in the polytope $H_d$, its coordinates satisfy
\begin{equation}
  \mathbf{v}_{1,\sigma(\{i\})}+\mathbf{v}_{2,\sigma(\{i\})}=1, \forall i\in[n].
\end{equation}
\end{lemma}

Based on this result, Theorem~\ref{thm:departial} and Lemma~\ref{lm:cptvertex}, we can show that the DOP is indeed a combinatorial optimization problem over a set system.
\begin{theorem}\label{thm:dop}
\emph{(DOP is combinatorial optimization)} The defender oracle problem is, for any vector $\mathbf{w}\in\mathbb{R}^{|S|}$, maximize a pseudo-boolean function over a set system $\varepsilon$.
\begin{equation}\label{eq:01opt}
	\max\limits_{\mathbf{x}\in\varepsilon}\left[\sum\limits_{V\in S}\mathbf{w}_{\sigma(V)}\left(\prod\limits_{\{i\}\in V} \mathbf{x}_i\right)\right],
\end{equation}
\end{theorem}

Clearly, the complexity of the DOP is not only dependent on the set system, but also dependent on the support set $S$, which describes the degree of the above pseudo-boolean function. For example, If the attacker can attack at most two targets and the utility functions are non-additive, the support set $S=\{A\in 2^{[n]}||A|\leq2\}$. In this case, the DOP is a general constrained binary quadratic programming problem, which is NP-hard. In Section 5, we assume that all the utility functions are additive. In this case, Lemma~\ref{lm:add} shows that the support set $S=[n]$ and the defender oracle problem will degenerate to the following linear optimization problem:
\begin{equation}\label{eq:exampledop}
	\max\limits_{\mathbf{x}\in\varepsilon}\mathbf{w}^T\mathbf{x}.
\end{equation}
In this case, the complexity of such an oracle problem is only dependent on the complexity of set system $\varepsilon$. For example, if the defender can attack at most $k$ targets, the set system $\varepsilon$ is a uniform matroid and solving the DOP only requires summing first $k$ largest elements of $\mathbf{w}$. If the defender's resources are obtained at some costs and there exists a resource budget, the set system in this scenario encodes the knapsack problem, which is NP-hard (we can solve it in pseudo-polynomial time by dynamic programming).  More interesting applications can be found in~\cite{xu2016mysteries}. 

\begin{remark}
 Compared with the original complicated security game, the simple form of DOP not only provides a simpler method to design the polynomial time algorithm, but also facilities us the better understanding of the inherent complexity property of the security game. For example, in the non-zero sum and additive security game, the standard solution technique used is the Lemke's algorithm~\cite{lemke1964equilibrium}. Such an algorithm may run in exponential time and cannot provide any insights of the problem. Instead, our equivalence theorem reduces the complexity of the original problem into the complexity of the defender's pure strategy space, which   is dependent on the specific problem structure. Besides, In Section 6.2, we will also show how to utilize the DOP to solve several open problems in the security game domain.
\end{remark}

\begin{remark}
The previous results in~\cite{xu2016mysteries} demonstrates that the security game under single attacker resources and additive utility function is equivalent to the above oracle problem (\ref{eq:exampledop}). Their reduction is based on the equivalence between the optimization and the membership problem, in which the polynomial time solvability is dependent on the description length of the set system $\varepsilon$. However, the special case (\ref{eq:exampledop}) of our result strengthens their arguments, which is independent of the set system $\varepsilon$.
\end{remark}

\subsection{Applications}

We now provide some applications to illustrate the value of our theoretical development. The first direction is to utilize the results in Theorem~\ref{thm:dopandnonzero} to tackle several open problems on the complexity of the security game with multiple attacker resources and heterogenous defender resources~\cite{korzhyk2011security,xu2016mysteries}. Another direction is to answer the question in~\cite{xu2016mysteries}: what is the complexity of the security game when we consider both non-additive utility functions and multiple attacker resources.

Regarding the first direction, we provide a general algorithmic framework shown below (Algorithm~\ref{alg:nzzadd}). 

\begin{algorithm}[htb]
\renewcommand{\algorithmicrequire}{ \textbf{Input:}} 
\renewcommand{\algorithmicensure}{ \textbf{Output:}} 
\caption{General Framework for Non-zero-sum and Additive Security Game}   
\label{alg:nzzadd} 
\small 
\begin{algorithmic}  
\STATE 1. \textbf{Solve the compact problem:} Solve the linear program (\ref{eq:compactzero}) to obtain the optimal compact strategy $\mathbf{t}^*$ by ellipsoid method: (i) check the inequality constraint by exactly solving the polynomial-sized linear program (\ref{eq:cptzeroineq}); (ii) solve the membership problem by exactly solving the DOP (\ref{eq:exampledop}).
\STATE 2. \textbf{Convex decomposition:} Decompose optimal compact strategy $\mathbf{t}^*$ into the convex combination: $\mathbf{t}^*=\sum\limits_{i=1}^{n+1}\lambda_i \mathbf{v}^i$ by exactly solving the DOP (\ref{eq:exampledop}).
\STATE 3. \textbf{Vertex mapping:} Map each vertex $\mathbf{v}^i$ to a defender pure strategy $D_i$ by Algorithm~\ref{alg1}, output the defender's NE strategy: 

$\quad$ play pure strategy $D_i$ with probability $\lambda_i$, $1\leq i\leq n+1$.
\end{algorithmic}
\end{algorithm}

The framework outlined in Algorithm~\ref{alg:nzzadd} provides a possible path to design a polynomial time algorithm to solve the NE\footnote{The algorithm to solve the SSE is similar.} of the non-zero-sum and additive security game. The details in the second step of Algorithm~\ref{alg:nzzadd} can be found in~\cite{grotschel1981ellipsoid}. The key lies in the polynomial time solvability of the DOP (\ref{eq:exampledop}), which is completely dependent on the combinatorial problem encoding by the set system $\varepsilon$.

\textbf{LAX airport checkpoint placement problem}~\cite{pita2008deployed}. This problem is one of the earliest applications of security games. In this setting, the security force has $k$ police officers that are to be deployed across $n$ (where $k < n$) checkpoints.Each police officer can be deployed at any given check point. Therefore, any subset of $[n]$ of size at most $k$ is a defender pure strategy. \cite{korzhyk2011security} extends this game model into the multiple attacker resources and shows that this problem can still be solved in poly$(n)$ time by a state transition algorithm~\cite{korzhyk2011security}. In our framework, the DOP is the linear optimization over a uniform matroid.
\begin{align}
&\max\quad \mathbf{w}^T\mathbf{x}\\
& \begin{array}{r@{\quad}l@{}l@{\quad}l}
s.t. &\sum\limits_{i=1}^{n}\mathbf{x}_i\leq k, \mathbf{x}\in\{0,1\}^n.\qquad \notag\\
\end{array} 
\end{align}
The above problem can be solved in polynomial time by summing the $k$ largest elements of vector $\mathbf{w}$. Thus, it verifies previous results.

In the following three cases, the defender's resources are heterogenous such that there exists some practical constrains in the set system $\varepsilon$. 

\textbf{Geographic constrained patrolling problem}. In the patrolling problem, due to geographic constraints, the police officer can only patrol the area around the station. In this case, the resources of different defenders (police) can defend different groups of targets. In our framework, we can construct a weighted bipartite graph as follows: (i) two disjoint sets $U,V$, $U$ represents all the nodes, $V$ represents all the resources; (ii) there exists an edge between the node $u$ in $U$ and node $v$ in $V$ if the resource $v$ can cover node $u$; (iii) associate each edge (u,v) with a weight $\mathbf{w}_u$ ($\mathbf{w}$ is the vector in the DOP). Then the DOP is a weighted bipartite matching problem, which can be solved in polynomial time by Hungarian algorithm. 
 
\textbf{Federal air marshal scheduling problem}~\cite{tsai2009iris}. In such applications, one air marshal is assigned to protect several sequential flights with the constraint that any destination of the previous flight is the departure of the next flight. The objective is to cover all current flights.  \cite{korzhyk2010complexity} investigates this problem under single attacker resources and shows the polynomial solvability in some cases and NP-hardness in other cases. However, attackers may initiate simultaneous attacks (e.g., the flights of 911) and there still does not exist any efficient algorithm. In our framework, we can construct the following weighted set cover problem: let the node set $[n]$ be the universe and all the air marshals constitute the collection $S$ of subsets of $[n]$; then associate the weight $\mathbf{w}$ to each element of the universe. Then, the DOP is a weighted set cover problem and our results show that when the attacker has multiple resources, the problem is generally NP-hard but we can still solve this problem in some cases. For example, if each air marshal can protect at most two flights (a pair of round trip flights), the set system $\varepsilon$ indeed encodes the weighted $2-$cover, which can be solved in poly($n$) time.

\textbf{Spatio-temporal security game}~\cite{fang2013optimal,fang2016deploying}. Among many applications of security games, an important class is the spatio-temporal security game. This kind of game is used to model the games played in the spatio-temporal spaces such as scheduling patrol boats of the US Coast Guard~\cite{fang2013optimal}, wildlife protection~\cite{fang2016deploying}. The current solution technique of this game is to discretize the space and time and build $2-$D gird, in which the security force patrol the points. Combining the results in~\cite{xu2014solving}, we can show that spatio-temporal security game with multiple attacker resources are indeed a min-cost flow problem, which can be solved in poly$(n)$ time.

There exists other applications that can be cast in our framework such as passenger screening for the Transportation Security Administration~\cite{brown2016one}. Indeed, based on our general framework in Algorithm~\ref{alg:nzzadd}, all the results under the single attacker resources can be directly extended to the scenario of multiple attacker resources.

Now, we focus on the second direction discussed earlier to answer the question on the complexity of the security game with non-additive utility functions and multple resources. 

\begin{corollary}
  The security game with general non-additive utility functions are NP-hard.
\end{corollary}
The answer appears to be negative since the constrained binary optimization problem is always NP-hard. However, inspired by the work~\cite{wang2016non}, we can still explore some interesting polynomial solvable classes under the following standard relaxation technique: \emph{replace the hard constraints on defender's pure strategy by the ``soft'' cost function}, in which defender's pure strategy space $\mathcal{D}=2^{[n]}$ but each strategy $D\in\mathcal{D}$ is associated with an additive cost function. More specifically, when the attacker and defender choose strategy $A$ and $D$, respectively, the attacker's and defender's pay off is given by $B_a(A\backslash D)+L_a(A\cap D)-C_a(A)$ and $L_d(A\backslash D)+B_d(A\cap D)-C_d(D)$, where $C_a(\cdot)$ and $C_d(\cdot)$ is the attacker's and defender's cost function, respectively. In this case, we can still apply our framework to reduce the equilibrium computation to the above oracle problem.
\begin{corollary}
There is a \emph{poly}($n$) time algorithm to compute the defender's Nash equilibrium in the non-additive security game with cost function, \textbf{if and only if} there is a \emph{poly}($n$) time algorithm to compute the the following problem: for any given $\mathbf{w}\in\mathbb{R}^{|S|}$
\begin{equation}\label{eq:01opt}
	\max\limits_{\mathbf{x}}\left[\sum\limits_{V\in S}\mathbf{w}_{\sigma(V)}\left(\prod\limits_{\{i\}\in V} \mathbf{x}_i\right)\right],
\end{equation}
\end{corollary}
 Such a problem is an unconstrained optimization problem and we can apply the known results in the combinatorial algorithm design to obtain some polynomial solvable classes. The proof of the above corollary is quite similar with our previous theoretical development, and the only difference is that we require decomposition of the matrix involving cost functions (see the Theorem 1 in~\cite{wang2016non}). Then, based on the above result, we can design a similar framework as in Algorithm~\ref{alg:nzzadd} to solve the zero-sum and non-additive security game and develop the following two applications. The detailed technical proof can be found in~\cite{wang2016non}.
 
\textbf{The security game in a tree network}. A sensor network is often in the form of a tree topology. The game is such that the attacker attempts to invade some nodes to destroy the connectedness of the network and the IT manager is required to deploy the anti-virus software in some nodes. We can show that the DOP of this game is a sub-modular minimization problem, which can be solved in poly$(n)$ time.

\textbf{The separable support set $S$}. Here the ``separable'' is defined as, $S= \bigcup_{i=1}^{m} S_i$ such that $A_i\cap A_j=\emptyset, \forall A_i\in S_i, A_j\in S_j, i\neq j$, the component is defined as $U_i=\cup_{U\in S_i} U$ with $\max_i |U_i|=\Theta(\log(n))$. In this case, we can solve the DOP in poly$(n)$ time via an enumerating algorithm. This result can be applied to a sparse network. For example, if the network is sparse enough such that the size of largest connected component of the network is $\Theta(\log(n))$, support $S$ will be separable in this case.

\section{Conclusions and Future Work}

In this paper, we examined the security game under multiple attacker resources and non-additive utility functions. In particular, we utilized the transformation and projection of the polytope to compactly represent the game in polynomial size. Further, based on the geometric structure of the projected polytope, we designed two poly$(n)$ time vertex mapping algorithm and showed that the equilibrium computation problem is equivalent to a defender oracle problem. Finally, we proved that such a defender oracle problem is essentially a combinatorial optimization problem over a set system. This result not only resolves some open problems in the security game domain, but also provides a new path to explore the polynomial solvable classes in more complex scenarios. Our proposed theoretical framework is significantly different from the state-of-the-art techniques used in the security game domain and greatly extends both the NP-hard and polynomial solvable classes. 

For future work, we plan to investigate the following direction:  
\begin{itemize}
  \item  \textbf{The computation of Nash equilibrium of non-zero-sum and non-additive security game}. We have utilized a constructed linear transformation to convert the non-zero-sum and additive security game to a $n-$dimensional saddle point problem.  Instead, when the utility function is non-additive, the construction of such a transformation is still open.
  \item \textbf{The approximate version of our equivalence theorem}. It is interesting to understand how an approximate defender oracle relates to the approximate equilibrium computation, since the problem is easy to falls into the category of NP-hard when we consider non-additive utility functions. One possible path is to exploit the no-regret learning framework proposed in~\cite{immorlica2011dueling}.
\end{itemize}



%
\bibliographystyle{unsrt}
\bibliography{sigproc}  
%
%
\appendix
\section{Lemmas in Section 3}
\subsection{Proof of Lemma~\ref{lm:com}}
\begin{proof}
 Indeed, we can use the existing results in m$\ddot{o}$bius and zeta transform to show the correctness of such a relationship. Here we provide a different proof, which is based on the method of counting in another way. Here we only prove for the attacker's benefit function, the proof for the other utility functions is similar.
 \begin{align}
		&\sum\limits_{V\subseteq U}B^c_a(V)\mathop=\limits^{(a)}\sum\limits_{V\subseteq U}\sum\limits_{W\subseteq V}(-1)^{|V\backslash W|} B_a(W)\notag\\
		&\mathop=\limits^{(b)}\sum\limits_{V\subseteq U}\sum\limits_{j=0}^{|V|}(-1)^{|V|-j}\cdot\sum\limits_{W\subseteq V,|W|=j}B_a(W)\notag\\
		&\mathop=\limits^{(c)}\sum\limits_{i=0}^{|U|}\sum\limits_{V\subseteq U,|V|=i}\sum\limits_{j=0}^{i}(-1)^{i-j}\cdot\sum\limits_{W\subseteq V,|W|=j}B_a(W)\label{eq:pf:decom1}\\
		&\mathop=\limits^{(d)}\sum\limits_{j=0}^{|U|}\sum\limits_{|W|=j,W\subseteq U}\left[\sum\limits_{i=j}^{|U|}\binom{|U|-j}{i-j}(-1)^{i-j}\right]\cdot B_a(W)\label{eq:pf:decom2}\\
		&\mathop=\limits^{(e)}B_a(U).\notag
\end{align}
(a) above is based on the definition of common utilities, (b) and (c) rearrange all the summation terms via their cardinality, (d) utilizes the following trick: we first traverse all the subsets of $U$ and then for each subset $V$, traverse its subset $W$. Thus, there are multiple copies for a specific $B(W)$ of Equation (\ref{eq:pf:decom1}). Equation (\ref{eq:pf:decom2}) derives from counting the coefficient of $B_a(W)$ in a backward way. We first fix set $W$ and suppose $|W|=j$, and the number of set $V$ with $|V|=i$ containing $W$ as the subset is $\binom{|U|-j}{i-j}$. The corresponding coefficient of $B_a(W)$ is $(-1)^{i-j}$. Then traversing all the possible set $V$, i.e., traverse at the order of cardinality, we have the coefficient of $B_a(W)$ is
	\begin{align}
		\sum\limits_{i=j}^{|U|}\binom{|U|-j}{i-j}(-1)^{i-j}=\sum\limits_{i=0}^{|U|-j}\binom{|U|-j}{i}(-1)^{i}.
	\end{align}
The above coefficient is equal to $1$ iff $j=|U|$, and $0$, otherwise.  Thus, the lemma follows.
\end{proof}

\subsection{Proof of Lemma~\ref{lm:decomplete}}
\begin{proof}
We first show that the matrix $\mathbf{L}^A$ can be transformed into a similar structure of matrix $\mathbf{B}^A$. Define 
\begin{equation}
  \widetilde{\mathbf{L}}^A\mathbf{P}=\mathbf{L}^A,
\end{equation}
where $ \widetilde{\mathbf{L}}^A_{\sigma(A),\mu(D)}=L_a(A\cap D^c)$.
Based on the definition of matrix $\mathbf{P}$, for arbitrary $A,D\in 2^{[n]}$, we have
\begin{align}
 \mathbf{L}^A_{\sigma(A),\mu(D)}&=\sum\limits_{U\in 2^{[n]}}\widetilde{\mathbf{L}}^A_{\sigma(A),\mu(U)}\mathbf{P}_{\mu(U),\mu(D)}\notag\\
  &\mathop=\limits^{(a)} \sum\limits_{U\in 2^{[n]}}\widetilde{\mathbf{L}}^A_{\sigma(A),\mu(U)}\mathbf{P}_{\sigma(U^c),\mu(D)}\notag\\
  &=\sum\limits_{U\in 2^{[n]}} L_a(A\cap U^c) \cdot \mathbbm{1}\{U^c=D\}\notag\\
  &= L_a(A\cap D).\notag
\end{align}
(a) is based on our choice of index function $\mu(U^c)=\sigma(U)$. Then we can observe that the matrix $\widetilde{\mathbf{L}}^A$ and $\mathbf{B}^A$ has a similar form. Thus we only need to prove the decomposition of matrix $\mathbf{B}^A$, and the other one follows.

For arbitrary $A,D\in 2^{[n]}$, the element $(\sigma(A),\mu(D))$ of matrix $\mathbf{Q}\mathbf{D}^B\mathbf{Q}^T$ is equal to
	\begin{align}
		&\mathbf{Q}_{\sigma(A)}\mathbf{D}^B\mathbf{Q}_{\mu(D)}^T.
	\end{align}
	Let vector $\mathbf{T}=\mathbf{Q}_{\sigma(A)}\mathbf{D}^B$, Based on the definition of matrix $\mathbf{Q}$ and $\mathbf{D}^B$, the $\mu(U)$th coordinate of vector $\mathbf{T}$ is
	\begin{align}
		\mathbf{T}_{\mu(U)}&=\sum\limits_{V\in2^{[n]}}\mathbf{Q}_{\sigma(A),\mu(V)}\cdot\mathbf{D}_{\mu(V),\mu(U)}^B\notag\\
		&\mathop=\limits^{(a)}\sum\limits_{V\in2^{[n]}}\mathbf{Q}_{\sigma(A),\mu(V)}\cdot\mathbf{D}_{\sigma(V^c),\mu(U)}^B \notag\\
		&\mathop=\limits^{(b)}\sum\limits_{V\in2^{[n]}}\mathbbm{1}\{V^c\subseteq A\} \cdot \mathbbm{1}\{V=U\}\cdot B^c_a(U^c)\notag\\
		&=\mathbbm{1}\{U^c\subseteq A\}\cdot B^c(U^c).
	\end{align}
(a) above is based on our choice of index function $\mu(U)=\sigma(U^c)$ for arbitrary $U\in2^{[n]}$, (b) is based on the definition of matrix $\mathbf{Q}$ and $\mathbf{D}^B$.
	Then, we have
	\begin{align}
		\mathbf{Q}_{\sigma(A)}\mathbf{D}^B\mathbf{Q}_{\mu(D)}^T&=\sum\limits_{U\in2^{[n]}}\mathbf{T}_{\mu(U)}\cdot\mathbf{Q}_{\mu(U),\mu(D)}^T\notag\\
		&\mathop=\limits^{(a)}\sum\limits_{U\in2^{[n]}}\mathbf{T}_{\mu(U)}\cdot\mathbf{Q}_{\sigma(D^c),\mu(U)}\notag\\
		&=\sum\limits_{U\in2^{[n]}}\mathbbm{1}\{U^c \subseteq A\}\cdot \mathbbm{1}\{U^c \subseteq D^c\}\cdot B^{c}(U^c)\notag\\
		&=\sum\limits_{U\subseteq A\cap D^c}B^{c}_a(U)\notag\\
		&\mathop=\limits^{(b)}B_a(A\cap D^c).
	\end{align}
(a) above is based on our choice of index function $\mu(U)=\sigma(U^c)$ for arbitrary $U\in2^{[n]}$, (b) utilizes Lemma~\ref{lm:com}.  Since $A$ and $D$ is arbitrarily chosen, based on the definition of the benefit matrix $\mathbf{B}^A$, we have
\begin{equation}
  \mathbf{B}^A=\mathbf{Q}\mathbf{D}^B\mathbf{Q}^T.
\end{equation}
Similarly, we have
\begin{equation}
 \mathbf{L}^A=\widetilde{\mathbf{L}}^A\mathbf{P}=\mathbf{Q}\mathbf{D}^L\mathbf{Q}^T\mathbf{P}.
\end{equation}
Thus, the lemma follows.
\end{proof}

\subsection{The Proof of Lemma~\ref{lm:completeandpartial}}
\begin{proof}
 We only prove for the benefit matrix $\mathbf{B}^{a}$, the proof is similar for loss matrix $\mathbf{L}^{a}$.
 
For any $A\in\mathcal{A}, D\in\mathcal{D}$, the element $(\sigma(A),\mu(D))$ of matrix $\mathbf{S}\mathbf{B}^A\mathbf{R}$ is equal to
	\begin{align}
		&\mathbf{S}_{\sigma(A)}\mathbf{B}^A\mathbf{R}_{\mu(D)}.
	\end{align}

Let vector $\mathbf{T}=\mathbf{S}_{\sigma(A)}\mathbf{B}^A$, The $\mu(U)$th coordinate of vector $\mathbf{T}$ is
\begin{align}
  \mathbf{T}_{\mu(U)}&=\sum\limits_{V\in 2^{[n]}}\mathbf{S}_{\sigma(A),\mu(V)}\cdot\mathbf{B}^A_{\mu(V),\mu(U)}\notag\\
  &\mathop=\limits^{(a)}\sum\limits_{V\in 2^{[n]}}\mathbf{S}_{\sigma(A),\mu(V)}\cdot\mathbf{B}^A_{\sigma(V^c),\mu(U)}\notag\\
  &\mathop=\limits^{(b)}\sum\limits_{V\in 2^{[n]}} \mathbbm{1}\{A=V^c\}\cdot B_{a}(V^c\cap U^c)\notag\\
  &=B_{a}(A\cap U^c)\notag.
\end{align}
(a) above is based on our choice of index function, (b) is based on the definition matrix $\mathbf{S}$ and $\mathbf{B}^A$. Then we have
\begin{align}
\mathbf{S}_{\sigma(A)}\mathbf{B}^A\mathbf{R}_{\mu(D)}&=\sum\limits_{U\in 2^{[n]}} \mathbf{T}_{\mu(U)}\cdot\mathbf{R}_{\mu(U), \mu(D)} \notag \\
&\mathop=\limits^{(a)}\sum\limits_{U\in 2^{[n]}} \mathbf{T}_{\mu(U)}\cdot\mathbf{R}_{\sigma(U^c), \mu(D)} \notag \\
&\mathop=\limits^{(a)}\sum\limits_{U\in 2^{[n]}} B_{a}(A\cap U^c) \cdot \mathbbm{1}\{D=U\}\notag\\
&=B_{a}(A\cap D^c)\notag.
\end{align}
(a) above is based on our choice of index function, (b) is based on the definition of matrix $\mathbf{R}$. Since $A$ and $D$ is arbitrarily chosen from attacker's and defender's pure strategy space, based on the definition of matrix $\mathbf{B}^a$, we have
\begin{equation}
  \mathbf{B}^a=\mathbf{S}\mathbf{B}^A\mathbf{R}.
\end{equation}
Thus, the lemma follows.
\end{proof}

\subsection{The Proof of Theorem~\ref{thm:departial}}
\begin{proof}
  Based on the results of Lemma~\ref{lm:decomplete} and Lemma~\ref{lm:completeandpartial}, we have
  \begin{equation}
    \mathbf{B}^a=\mathbf{S}\mathbf{Q}\mathbf{D}^B\mathbf{Q}^T\mathbf{R}.
  \end{equation}
Considering the fact that the matrix $\mathbf{D}^B$ is diagonal, we can further write it as
 \begin{equation}
    \mathbf{B}^a=\mathbf{S}\mathbf{Q}\mathbf{S}\mathbf{D}^b\mathbf{S}\mathbf{Q}^T\mathbf{R}.
  \end{equation}
  Let matrix $\mathbf{E}=\mathbf{S}\mathbf{Q}\mathbf{S}$ and $\mathbf{J}=\mathbf{S}\mathbf{Q}^T\mathbf{R}$. Using a similar argument to the proof of Lemma~\ref{lm:completeandpartial}, we have
 \begin{align}
   &\mathbf{E}_{\sigma(A),\sigma(U)}=\mathbbm{1}\{U\subseteq A\}, \forall A,U\in \mathcal{A},\notag\\
   &\mathbf{J}_{\sigma(A),\mu(D)}=\mathbbm{1}\{A\subseteq D^c\}, \forall A\in \mathcal{A}, D\in \mathcal{D}.\notag
 \end{align}
 Similarly, we have $\mathbf{K}=\mathbf{S}\mathbf{Q}^T\mathbf{P}\mathbf{R}$ and 
 \begin{equation*}
 \mathbf{K}_{\sigma(A),\mu(D)} =\mathbbm{1}\{A\subseteq D\}, \forall A\in \mathcal{A}, D\in \mathcal{D}.
 \end{equation*}
 The proof for loss matrix $\mathbf{L}^a$ is quite similar. Thus, the theorem follows. 
\end{proof}

\subsection{The Proof of Theorem~\ref{thm:cpt}}

\begin{proof}
  The correctness of our compactly represented model is based on the following argument:
\begin{align*}
	&(\mathbf{p}^*, \mathbf{q}^*)\text{ is a NE of the zero-sum security game} \\
	\Rightarrow &\quad (\mathbf{p}^*)^T\mathbf{M}^{a}\mathbf{q^*}\geq \mathbf{p}^T\mathbf{M}^{a}\mathbf{q^*}, \forall \mathbf{p}\in \Delta_{N_a} \\
	\Rightarrow &\quad [f(\mathbf{p}^*)]^T(\mathbf{D}^bg_1(\mathbf{q^*})+\mathbf{D}^bg_2(\mathbf{q^*}))\geq \\
	&\qquad\qquad [f(\mathbf{p})]^T(\mathbf{D}^bg_1(\mathbf{q^*})+\mathbf{D}^bg_2(\mathbf{q^*})), \forall \mathbf{p}\in \Delta_{N_a}\\
	\Rightarrow &\quad [\pi_{S}(f(\mathbf{p}^*))]^T(\widetilde{\mathbf{D}}^b\pi_{S}(g_1(\mathbf{q}^*))+\widetilde{\mathbf{D}}^l\pi_{S}(g_2(\mathbf{q}^*)))\geq\\
	&\qquad\qquad \bar{\mathbf{p}}^T(\widetilde{\mathbf{D}}^b\pi_{S}(g_1(\mathbf{q}^*))+\widetilde{\mathbf{D}}^l\pi_{S}(g_2(\mathbf{q}^*))),\forall \bar{\mathbf{p}}\in H_a\\
	\Rightarrow &\quad \pi_S(f(\mathbf{p}^*))\text{ is the optimal solution of (\ref{eq:cptmodel1}).} 
\end{align*}
Similarly, we can obtain that $(\pi_S(g_1(\mathbf{q}^*)),\pi_S(g_2(\mathbf{q}^*)))$ is the optimal solution of (\ref{eq:cptmodel1}). The reverse direction is based on the following argument: given ($\mathbf{p}^*,\mathbf{q}^*$), 
\begin{align*}
	&(\pi_S(f(\mathbf{p}^*)),(\pi_S(g_1(\mathbf{q}^*)),\pi_S(g_2(\mathbf{q}^*))))\text{ is the optimal solution } \\
	&\text{of (\ref{eq:cptmodel1})}\\
	\Rightarrow & [\pi_{S}(f(\mathbf{p}^*))]^T(\widetilde{\mathbf{D}}^b\pi_{S}(g_1(\mathbf{q}^*))+\widetilde{\mathbf{D}}^l\pi_{S}(g_2(\mathbf{q}^*)))\geq\\
	&\qquad\qquad \bar{\mathbf{p}}^T(\widetilde{\mathbf{D}}^b\pi_{S}(g_1(\mathbf{q}^*))+\widetilde{\mathbf{D}}^l\pi_{S}(g_2(\mathbf{q}^*))),\forall \bar{\mathbf{p}}\in H_a\\
	\Rightarrow & [f(\mathbf{p}^*)]^T(\mathbf{D}^bg_1(\mathbf{q^*})+\mathbf{D}^bg_2(\mathbf{q^*}))\geq [f(\mathbf{p})]^T(\mathbf{D}^bg_1(\mathbf{q^*})+\\
	&\mathbf{D}^bg_2(\mathbf{q^*})), \forall  \mathbf{p}\in \{\mathbf{p}\in\Delta_{N_a}|\pi_S(f(\mathbf{p}))=\bar{\mathbf{p}}\},\bar{\mathbf{p}}\in H_a\\
	\Rightarrow &(\mathbf{p}^*)^T\mathbf{M}^{a}\mathbf{q^*}\geq \mathbf{p}^T\mathbf{M}^{a}\mathbf{q^*},\\
	&\forall  \mathbf{p}\in \{\mathbf{p}\in\Delta_{N_a}|\pi_S(f(\mathbf{p}))=\bar{\mathbf{p}}\},\bar{\mathbf{p}}\in H_a,\\
	\mathop\Rightarrow\limits^{(a)} &(\mathbf{p}^*)^T\mathbf{M}^{a}\mathbf{q^*}\geq \mathbf{p}^T\mathbf{M}^{a}\mathbf{q^*}, \forall \mathbf{p}\in \Delta_{N_a}\\
	\Rightarrow &\mathbf{p}^* \text{ is a NE of the zero-sum security game}. 
\end{align*}	
(a) is based on the fact that,
\begin{equation}
	\Delta_{N_a}=\bigcup\limits_{\bar{\mathbf{p}}\in H_a} \{\mathbf{p}\in\Delta_N|\pi_S(f(\mathbf{p}))=\bar{\mathbf{p}}\}.
\end{equation}
Clearly, we have the following direction, 
\begin{equation}
	\bigcup\limits_{\bar{\mathbf{p}}\in H_a} \{\mathbf{p}\in\Delta_{N_a}|\pi_S(f(\mathbf{p}))=\bar{\mathbf{p}}\}\subseteq \Delta_{N_a}.
\end{equation}
The other direction is based on,
\begin{align*}
	\forall \mathbf{x}\in \Delta_{N_a} &\Rightarrow f(\mathbf{x})\in \Delta_{N_a}^a\\
	&\Rightarrow \pi_S(f(\mathbf{x}))\in H_a\\
	&\Rightarrow \exists \bar{\mathbf{x}}\in H_a \text{ s.t. } \pi_S(f(\mathbf{x}))=\bar{\mathbf{x}}\\
	&\Rightarrow \mathbf{x} \in \{\mathbf{p}\in\Delta_N|\pi_S(f(\mathbf{p}))=\bar{\mathbf{x}}\}\\
	&\Rightarrow \mathbf{x} \in \bigcup\limits_{\bar{\mathbf{p}}\in H_a} \{\mathbf{p}\in\Delta_N|\pi_S(f(\mathbf{p}))=\bar{\mathbf{p}}\}.
\end{align*}
Similarly, we can obtain that $\mathbf{q}^*$ is a NE of the NASG. Thus, theorem follows.
\end{proof}

\subsection{The Proof of Lemma~\ref{lm:convex}}
\begin{proof}
 The polytope $H_a$ is convex follows directly from the fact the that the simplex $\Delta_{N_a}$ is convex and the linear transformation and projection of a convex polytope is also a convex polytope. To show the convexity of the polytope $H_d$, consider any two points: $(\mathbf{x}_1,\mathbf{x}_2), (\mathbf{y}_1,\mathbf{y}_2)\in H_d$, where
 \begin{align}
   &\mathbf{x}_1=\pi_S(\mathbf{J}\mathbf{q}_1), \mathbf{x}_2=\pi_S(\mathbf{K}\mathbf{q}_1), \mathbf{q}_1\in \Delta_{N_d},\notag\\
   &\mathbf{y}_1=\pi_S(\mathbf{J}\mathbf{q}_2), \mathbf{y}_2=\pi_S(\mathbf{K}\mathbf{q}_2), \mathbf{q}_2\in \Delta_{N_d}.\notag
 \end{align}
 Then we have the following convex combination
 \begin{align}
   &\lambda (\mathbf{x}_1,\mathbf{x}_2) +(1-\lambda) (\mathbf{y}_1,\mathbf{y}_2)=(\lambda\mathbf{x}_1+(1-\lambda)\mathbf{y}_1,  \lambda\mathbf{x}_2+(1-\lambda)\mathbf{y}_2)\notag\\
   &=(\lambda\pi_S(\mathbf{J}\mathbf{q}_1)+(1-\lambda)\pi_S(\mathbf{J}\mathbf{q}_2),  \lambda\pi_S(\mathbf{K}\mathbf{q}_1)+(1-\lambda)\pi_S(\mathbf{K}\mathbf{q}_2))\notag\\
   &\mathop=\limits^{(a)} (\pi_S(\mathbf{J}(\lambda\mathbf{q}_1+(1-\lambda)\mathbf{q}_2)), \pi_S(\mathbf{K}(\lambda\mathbf{q}_1+(1-\lambda)\mathbf{q}_2)))\notag\\
   &\in H_d.\notag
 \end{align}
(a) above is based on the fact that the projection operator is linear, the last step based on the convexity of $\Delta_{N_d}$ such that $\lambda\mathbf{q}_1+(1-\lambda)\mathbf{q}_2\in \Delta_{N_d}$.
\end{proof}

\section{Lemmas in Section 4}

\subsection{Proof of Lemma~\ref{lm:miop}}

\begin{proof}
 We only need to show that the vertices of $H_a$ are the rows of the sub-matrix of $\mathbf{E}$. The proof is similar with the proof of Lemma~\ref{lm:cptvertex}, and details can be seen in Appendix~\ref{app:cptvertex}.

The set $\{\pi_{S}(\mathbf{E}_{\sigma(A)})| \forall A\in \mathcal{A}\}$ contains all the vertices in $H^a$, where $\mathbf{E}_{\sigma(A)}$ is the $\sigma(A)$th row of matrix $\mathbf{A}$. We show that, for arbitrary pure strategy $A$ and support set $S$, $\mathbf{v}^{A}=\pi_{S}(\mathbf{E}_{\sigma(A)})$ is indeed an vertex of polytope $H^a$. This result implies that the set $\{\pi_{S}(\mathbf{E}_{\sigma(A)})| \forall A\in\mathcal{A}\}$ is equal to $I_a$.

Define 
\begin{equation}
	I'\triangleq\{\mathbf{v}^{U}=\pi_{S}(\mathbf{E}_{\sigma(A)})|\forall U\in 2^{[n]}\}.
\end{equation}
	
For any point $\mathbf{v}^{A_0}$ in $I'$, let
	\begin{equation}
		\mathbf{c}_{\sigma(V)}=1-\mathbbm{1}\{V\subseteq A_0\},
	\end{equation}
based on the definition of matrix $\mathbf{E}$, we have 
	\begin{align*}
		\mathbf{c}^T\mathbf{v}^{A_0}&=\sum\limits_{V\in S}\mathbf{c}_{\sigma(V)}\mathbf{v}^{A_0}_{\sigma(V)}\\
		&=\sum\limits_{V\in S}(1-\mathbbm{1}\{V\subseteq A_0\})\cdot\mathbbm{1}\{V\subseteq A_0\}\\
		&=0.
	\end{align*}
Considering the fact that all the $\mathbf{v}^{A}$ only have binary coordinates, we have $\mathbf{c}^T\mathbf{v}^{A}\geq 0=\mathbf{c}^T\mathbf{v}^{A_0}$. Therefore, the point $\mathbf{v}^{A_0}$ is a vertex of $H_a$. Also, such vertices are the rows of a sub-matrix of $\mathbf{E}$, which is formed by deleting the columns whose index does not belong to $\sigma(S)$. Since the dimension of matrix $\mathbf{E}$ is poly$(n)$ and we have the close-form expression of the vertices in $H_a$, the inequality constraint problem can be implicitly checked.
\end{proof}

\subsection{Proof of Lemma~\ref{lm:zerosumdopoandcpto}}

\begin{proof}
One direction is trivial: considering an arbitrary $(\bar{\mathbf{q}}_1,$ $\bar{\mathbf{q}}_2),u$ for separation problem of compact problem, taking $(\bar{\mathbf{q}}_1, \bar{\mathbf{q}}_2),$ $u$ as the input of the separation problem for $H_d$, if the answer is no, then we can output a hyperplane separating $(\bar{\mathbf{q}}_1, \bar{\mathbf{q}}_2), u$ from $H_d$; if the answer is yes, we obtain that $(\bar{\mathbf{q}}_1, \bar{\mathbf{q}}_2), u$ belongs to $H_d$ and implicitly check if the inequality constraints are satisfied.  Thus, we can solve the separation problem for compact problem in poly$(n)$ time if the separation problem for $H_d$ can be solved in poly$(n)$ time. 

For the other direction, we first suppose that the separation problem for the compact problem can be solved in poly$(n)$ time. Then, we show how to reduce the separation problem for $H_d$ to the above separation problem. Consider an arbitrary input $(\bar{\mathbf{q}}_1,$ $\bar{\mathbf{q}}_2)$, let
\begin{equation}
	u_0=|S|(\|\widetilde{\mathbf{D}}^b\|+\|\widetilde{\mathbf{D}}^l\|)+2,
\end{equation}
where $\|\cdot\|$ is the spectral norm of a matrix. Then taking $(\bar{\mathbf{q}}_1,$ $\bar{\mathbf{q}}_2), u_0$ as the input of the separation problem of compact problem. If the answer is yes, we have $(\bar{\mathbf{q}}_1,$ $\bar{\mathbf{q}}_2)\in H_d$; if not, output a hyperplane such that
\begin{equation}\label{eq:pf:hplane}
	\mathbf{a}_1^T\bar{\mathbf{q}}_1+\mathbf{a}_2^T\bar{\mathbf{q}}_2+bu_0> \mathbf{a}_1^T\mathbf{q}_1+\mathbf{a}_2^T\mathbf{q}_2+bu,
\end{equation}
$\forall (\mathbf{q}_1,\mathbf{q}_2)\in H_d, \mathbf{v}^T(\widetilde{\mathbf{D}}^b\mathbf{q}_1+\widetilde{\mathbf{D}}^l\mathbf{q}_2)\leq u, \forall\mathbf{v}\in I_a$. Considering the fact that
\begin{align*}
	|\mathbf{v}^T(\widetilde{\mathbf{D}}^b\mathbf{q}_1&+\widetilde{\mathbf{D}}^l\mathbf{q}_2)|\mathop\leq\limits^{(a)}|\mathbf{v}^T\widetilde{\mathbf{D}}^b\mathbf{q}_1|+|\mathbf{v}^T\widetilde{\mathbf{D}}^l\mathbf{q}_2|\notag\\
&\mathop \leq\limits^{(b)} \|\mathbf{v}\|\cdot \|\widetilde{\mathbf{D}}^b\mathbf{q}_1\|+\|\mathbf{v}\|\cdot \|\widetilde{\mathbf{D}}^l\mathbf{q}_2\|\\
	&\mathop <\limits^{(c)}|S|^{\frac{1}{2}}\|\widetilde{\mathbf{D}}^b\||S|^{\frac{1}{2}}+|S|^{\frac{1}{2}}\|\widetilde{\mathbf{D}}^l\||S|^{\frac{1}{2}}\\
&=|S|(\|\widetilde{\mathbf{D}}^b\|+\|\widetilde{\mathbf{D}}^l\|)<u_0, \forall \mathbf{v}\in I_a, (\mathbf{q}_1,\mathbf{q}_2)\in H_d.
\end{align*}
Above, (a) is based on the triangle inequality, (b) is based on Cauchy - Schwarz inequality, (c) follows from the definition of spectral norm of a matrix, (d) utilizes the fact that all the vertices of $H_d$ is binary (details can be seen in Appendix). Let $u=u_0-1$ in (\ref{eq:pf:hplane}), we have,
\begin{equation}\label{thm:eq:reduction}
	\mathbf{a}_1^T\bar{\mathbf{q}}_1+\mathbf{a}_2^T\bar{\mathbf{q}}_2+bu_0> \mathbf{a}_1^T\mathbf{q}_1+\mathbf{a}_2^T\mathbf{q}_2+b(u_0-1),
\end{equation}
$\forall (\mathbf{q}_1,\mathbf{q}_2)\in H_d, \mathbf{v}^T(\widetilde{\mathbf{D}}^b\mathbf{q}_1+\widetilde{\mathbf{D}}^l\mathbf{q}_2)\leq u_0-1, \forall\mathbf{v}\in I_a$. However, the $(u_0-1)$ is large enough such that the following condition always holds true, 
\begin{equation}
	\mathbf{v}^T(\widetilde{\mathbf{D}}^b\mathbf{q}_1+\widetilde{\mathbf{D}}^l\mathbf{q}_2)\leq u_0-1, \forall\mathbf{v}\in I_a, \forall (\mathbf{q}_1,\mathbf{q}_2)\in H_d.
\end{equation}
Since $u$ is unbounded in (\ref{eq:pf:hplane}), the coefficient $b$ should be negative. Hence, (\ref{thm:eq:reduction}) implies that
\begin{equation}
	\mathbf{a}_1^T\bar{\mathbf{q}}_1+\mathbf{a}_2^T\bar{\mathbf{q}}_2> \mathbf{a}_1^T\mathbf{q}_1+\mathbf{a}_2^T\mathbf{q}_2, \forall (\mathbf{q}_1,\mathbf{q}_2)\in H_d.
\end{equation}
which is the desired separating hyperplane between $(\bar{\mathbf{q}}_1,$ $\bar{\mathbf{q}}_2)$ and $H_d$. Since the size of matrix $\widetilde{\mathbf{D}}^b$ and $\widetilde{\mathbf{D}}^l$ is $\Theta(n^c)$, the spectral norm $\|\widetilde{\mathbf{D}}^b\|=\|\widetilde{\mathbf{D}}^b\|=\text{poly}(n)$ and the size of $u_0$ is also poly$(n)$. Thus, the lemma follows. 
\end{proof}

\subsection{The Proof of Lemma~\ref{lm:cptvertex}\label{app:cptvertex}}
\begin{proof}
  Since the vertex in the projected polytope $H_d$ is an image of the vertex in the high-dimensional polytope $\Delta_{N_d}^d$, the total number of vertices in the projected polytope is less than $N_d$, and the set $\{\mathbf{v}^D=(\pi_{S}(\mathbf{J}_{\mu(D)}),\pi_S(\mathbf{K}_{\mu(D)}))| \forall D\in \mathcal{D}\}$ contains all the vertices in $H^d$. 
	
Define 
	\begin{equation}
	I'\triangleq\{\mathbf{v}^D=(\pi_{S}(\mathbf{J}_{\mu(D)}),\pi_S(\mathbf{K}_{\mu(D)}))| \forall D\in \mathcal{D}\}.
	\end{equation}
	Since $I_d\subseteq I'$, if any point $\mathbf{v}^{D_0}$ in $I'$ satisfies following condition,
\begin{equation*}
		\exists (\mathbf{c}_1, \mathbf{c}_2)\in \mathbb{R}^{2|S|} \text{ s.t. } \mathbf{c}_1^T\mathbf{v}^{D_0}_1+\mathbf{c}_2^T\mathbf{v}^{D_0}_2\leq \mathbf{c}^T_1\mathbf{v}^{D}_1+\mathbf{c}^T_2\mathbf{v}^{D}_2,
\end{equation*}
$\forall (\mathbf{v}^{D}_1,\mathbf{v}^{D}_2)\in I' \text{ and } (\mathbf{v}^{D}_1,\mathbf{v}^{D}_2)\neq (\mathbf{v}^{D_0}_1,\mathbf{v}^{D_0}_2)$
, it must be a vertex of polytope $H_d$. The reason derives from the definition of the vertex in a convex polytope and the fact that $I_d\subseteq I'$. For any point $\mathbf{v}^{D_0}$ in $I'$, based on the definition of matrix $\mathbf{J}$ and $\mathbf{K}$, let
\begin{equation*}
		\mathbf{c}_{1,\sigma(V)}=1-\mathbbm{1}\{V\subseteq D_0^c\},\text{ and } \mathbf{c}_{2,\sigma(V)}=1-\mathbbm{1}\{V\subseteq D_0\}.
\end{equation*}
	then we have 
	\begin{align*}
		\mathbf{c}_1^T\mathbf{v}^{D_0}_1+&\mathbf{c}_2^T\mathbf{v}^{D_0}_2=\sum\limits_{V\in S}\mathbf{c}_{1,\sigma(V)}^T\mathbf{v}^{D_0}_{1,\sigma(V)}+\mathbf{c}_{2,\sigma(V)}^T\mathbf{v}^{D_0}_{2,\sigma(V)}\\
		=&\sum\limits_{V\in S}(1-\mathbbm{1}\{V\subseteq D_0^c\})\cdot\mathbbm{1}\{V\subseteq D_0^c\}+\\
		&(1-\mathbbm{1}\{V\subseteq D_0\})\cdot\mathbbm{1}\{V\subseteq D_0\}\\
		=&0.
	\end{align*}
Considering the fact that all the points $(\mathbf{v}^{D}_1,\mathbf{v}^{D}_2)$ are binary vector, we have $\mathbf{c}^T_1\mathbf{v}^{D}_1+\mathbf{c}^T_2\mathbf{v}^{D}_2\geq 0=\mathbf{c}_1^T\mathbf{v}^{D_0}_1+\mathbf{c}_2^T\mathbf{v}^{D_0}_2$. Therefore, the point $\mathbf{v}^{D_0}$ is a vertex of $H_d$. And such vertices are the columns of a sub-matrix of $\begin{bmatrix}
\mathbf{J}\\
\mathbf{K}
\end{bmatrix}$, which is formed by deleting the row whose index does not belong to $\sigma(S)$. Thus, the lemma follows.
\end{proof}

\subsection{The Proof of Lemma~\ref{lm:cpt2org} and Lemma~\ref{lm:org2cpt}}
\begin{proof}
Our vertex mapping Algorithm~\ref{alg1} only enumerates all singleton subset $\{i\}$ and thus the running time is $O(n)$.  Considering an arbitrary pure strategy $D\in \mathcal{D}$ and the result of Lemma~\ref{lm:cptvertex}, we have
	\begin{align*}
		&\mathbf{v}^D_{1,\sigma(\{i\})}=1 \iff \{i\}\subseteq D^c,\\
		&\mathbf{v}^D_{2,\sigma(\{i\})}=1 \iff \{i\}\subseteq D.
	\end{align*}
	Thus, vertex mapping algorithm only needs to examine those singleton set to recover the subset $D$ and it correctly maps a vertex $\mathbf{v}^D$ in the projected polytope to a pure strategy $D$ in the original game. The vertex mapping  Algorithm~\ref{alg:pure2vertex} requires recovering each coordinate of the vertex in $H_d$, thus the running time is $O(|S|)=O(n^c)$. The correctness still follows the above argument.
\end{proof}

\section{The Lemmas in Section 5}

\subsection{The Proof of Lemma~\ref{lm:add}}
\begin{proof}
Here we only prove for the attacker's benefit function, and the proof for the other utility functions is similar. Based on the definition of the common utility, we have
\begin{align*}
  B_a^c(U)&=\sum\limits_{V\subseteq U}(-1)^{|U\backslash V|}B_a(V)\\
  &\mathop=\limits^{(a)}\sum\limits_{V\subseteq U}(-1)^{|U\backslash V|}\sum\limits_{i\in V}B_a(\{i\})\\
  &\mathop=\limits^{(b)}\sum\limits_{s=1}^{|U|}\sum\limits_{V\subseteq U, |V|=s}(-1)^{|U|-s}\sum\limits_{i\in V}B_a(\{i\})\\
  &\mathop=\limits^{(c)}\sum\limits_{i\in U} \sum\limits_{s=1}^{|U|} \binom{s-1}{|U|-1}(-1)^{|U|-s}\\
  &=\sum\limits_{i\in U}B_a(\{i\}) \sum\limits_{s=0}^{|U|-1} \binom{s}{|U|-1}(-1)^{|U|-1-s}\\
  &=\left\{\begin{matrix}
B_a(U), \text{ if } |U|=1\\ 
0, \text{ if } |U|>1
\end{matrix}\right..
\end{align*}
(a) above is based on the assumption that all the utility functions are additive, (b) rearranges all the summation terms via their cardinality, (c) utilizes the following trick: we first traverse all the subsets of $U$ and then for each subset $V$, traverse its element $i$. Thus, there are multiple copies for a specific $B_a(\{i\})$. Then we count the coefficient of $B_a(\{i\})$ in a backward way. We first fix set $\{i\}$ and suppose $|V|=s$, and the number of set $V$ with $|V|=s$ containing $i$ as its element is $\binom{|U|-1}{s-1}$. The corresponding coefficient of $B_a(\{i\})$ is $(-1)^{|U|-s}$. Then traversing all the possible set $V$, i.e., traverse at the order of cardinality, we arrive the desired formula. Thus, the lemma follows.
\end{proof}

\subsection{The Proof of Lemma~\ref{lm:zerosumsdd}}

\begin{proof}
It suffices to show the sufficiency of the following inequalities.
 \begin{align}
&U_a'(\mathbf{a}^*,\mathbf{t}^*)\geq  U_a'(\mathbf{a},\mathbf{t}^*) \text{ }\Leftarrow \text{ } U_a'(h(\mathbf{a}^*),\mathbf{t}^*)\geq U_a'(h(\mathbf{a}),\mathbf{t}^*)\notag\\
&U_d'(\mathbf{a}^*,\mathbf{t}^*)\geq  U_d'(\mathbf{a}^*,\mathbf{t}) \iff U_a'(h(\mathbf{a}^*),\mathbf{t}^*)\leq U_a'(h(\mathbf{a}^*),\mathbf{t}).\notag
 \end{align}
 
We first show the first one. One the one hand, we have
\begin{equation}
  U_a'(\mathbf{a}^*,\mathbf{t}^*)-U_a'(\mathbf{a},\mathbf{t}^*)=\sum\limits_{i=1}^{n}(a_i^*-a_i)[t_i^*L_d(i)+(1-t_i^*)L_d(i)].
\end{equation}
Similarly, on the other hand, we have
\begin{align}
U_a'(h(\mathbf{a}^*)&,\mathbf{t}^*)- U_a'(h(\mathbf{a}),\mathbf{t}^*)=\notag\\
&\sum\limits_{i=1}^{n}(h_i(a_i^*)-h_i(a_i))[t_i^*L_d(i)+(1-t_i^*)L_d(i)]
\end{align}
Given our transformation $h$, we have
\begin{align*}
  &U_a'(h(\mathbf{a}^*),\mathbf{t}^*)- U_a'(h(\mathbf{a}),\mathbf{t}^*) \geq 0\\
  \iff & \sum\limits_{i=1}^{n}(h_i(a_i^*)-h_i(a_i))[t_i^*L_d(i)+(1-t_i^*)L_d(i)] \geq 0\\
  \iff & \sum\limits_{i=1}^{n}\frac{B_d(i)-L_d(i)}{B_a(i)-L_a(i)}(a_i^*-a_i)[t_i^*L_d(i)+(1-t_i^*)L_d(i)] \geq 0\\
  \mathop\Rightarrow\limits^{(a)} & \sum\limits_{i=1}^{n}C(a_i^*-a_i)[t_i^*L_d(i)+(1-t_i^*)L_d(i)] \geq 0\\
  \Rightarrow& \sum\limits_{i=1}^{n}(a_i^*-a_i)[t_i^*L_d(i)+(1-t_i^*)L_d(i)] \geq 0.
\end{align*}
(a) above is based on the fact that 
\begin{equation}
 C=\max_{1\leq i\leq n}\frac{B_d(i)-L_d(i)}{B_a(i)-L_a(i)}\leq \infty,
\end{equation}
(b) is based on the fact that $B_d(i)>L_d(i)$ and $B_a(i)>L_a(i)$ (benefit is larger than loss).

Then we show the second one. One the one hand, we have
\begin{equation}
  U_d'(\mathbf{a}^*,\mathbf{t}^*)-U_d'(\mathbf{a}^*,\mathbf{t})=\sum\limits_{i=1}^{n}a_i^*(B_d(i)-L_d(i))(t^*_i-t_i).
\end{equation}
Similarly, on the other hand, we have
\begin{equation}
U_a'(h(\mathbf{a}^*),\mathbf{t})- U_a'(h(\mathbf{a}^*),\mathbf{t}^*)=\sum\limits_{i=1}^{n}h_i(a^*_i)(B_a(i)-L_a(i))(t_i^*-t_i)
\end{equation}
Given our transformation $h^{-1}$, we have
\begin{align*}
  &U_d'(\mathbf{a}^*,\mathbf{t}^*)-U_d'(\mathbf{a}^*,\mathbf{t}) \geq 0\\
  \iff & \sum\limits_{i=1}^{n}a_i^*(B_d(i)-L_d(i))(t^*_i-t_i) \geq 0\\
  \iff & \sum\limits_{i=1}^{n}h_i(a_i^*) \frac{B_a(i)-L_a(i)}{B_d(i)-L_d(i)}(B_d(i)-L_d(i))(t^*_i-t_i) \geq 0\\
  \iff & \sum\limits_{i=1}^{n}h_i(a_i^*) (B_a(i)-L_a(i))(t^*_i-t_i) \geq 0\\
  \iff & U_a'(h(\mathbf{a}^*),\mathbf{t})- U_a'(h(\mathbf{a}^*),\mathbf{t}^*) \geq 0.
\end{align*}
\end{proof}

\subsection{The Proof of Lemma~\ref{lm:hypercube}}
\begin{proof}
The proof can be regarded as the special case of the proof of Lemma~\ref{lm:miop}. In this case, the support set $S=[n]$, and we have any vertex $\mathbf{v}^{A}\in I_a$ is
\begin{equation}
 \mathbf{v}^{A}=\pi_{[n]}(\mathbf{E}_{\sigma(A)}), 
\end{equation}
and $\mathbf{v}^{A}_{\sigma(\{i\})}=\mathbbm{1}\{i\in A\}$.

According to the definition of the attacker's pure strategy space $\mathcal{A}=\{A\in 2^{[n]}||A|\leq c\}$, we have
\begin{equation}
  \sum\limits_{i=1}^{n}\mathbf{v}^{A}_{\sigma(\{i\})}=|A|\leq c.
\end{equation}
Since all the coordinates of the vertex $\mathbf{v}^{A}$ is binary, $I_d$ is equal to the intersection of the vertex set of a $n-$dimensional hypercube and hyperplane $\sum_{i=1}^{n}x_i\leq c$. Thus, the lemma follows. 
\end{proof}

\section{Lemmas in Section 6}

\subsection{The Proof of Lemma~\ref{lm:ncptvertex}}
\begin{proof}
  For an arbitrary vertex $(\mathbf{v}_1^D,\mathbf{v}_2^D)\in H_d$, based on the result of Lemma~\ref{lm:cptvertex}, we have
  \begin{equation}
   \mathbf{v}^D_{1,\sigma(\{i\})}=\mathbbm{1}\{i\in D^c\} \text{ and }  \mathbf{v}^D_{2,\sigma(\{i\})}=\mathbbm{1}\{i\in D\}.
 \end{equation}
 Then, we have
 \begin{align*}
 \mathbf{v}^D_{1,\sigma(\{i\})}+\mathbf{v}^D_{2,\sigma(\{i\})}&=\mathbbm{1}\{i\in D^c\}+\mathbbm{1}\{i\in D\}\\
 &=\mathbbm{1}\{i\in D^c\cup D\}\\
 &=\mathbbm{1}\{i\in 2^{[n]}\}=1.
 \end{align*}
Note that, since $\{i\}$ is the singleton set, we can write the $\{i\}\subseteq D$ as $i\in D$.
\end{proof}

\subsection{The Proof of Theorem~\ref{thm:dop}}
\begin{proof}
Based on the results of Lemma~\ref{lm:cptvertex}, for any vertex $(\mathbf{v}^{D}_1,$ $\mathbf{v}^{D}_2)\in I_d$, we have
\begin{equation*}
	\mathbf{v}^{D}_{1,\sigma(V)}=\mathbbm{1}\{V\subseteq D^c\}, \mathbf{v}^{D}_{2,\sigma(V)}=\mathbbm{1}\{V\subseteq D\}.
\end{equation*}
Then, we can write it as following equivalent form,
\begin{align*}
	\mathbf{v}^{D}_{2,\sigma(V)}&=\prod\limits_{i\in V}\mathbbm{1}\{\{i\}\in D\}=\prod\limits_{\{i\}\in V}\mathbf{v}^{D}_{2,\sigma(\{i\})}.
\end{align*}
\begin{align*}
	&\mathbf{v}^{D}_{1,\sigma(V)}=\prod\limits_{i\in V}\mathbbm{1}\{\{i\}\in D^c\}=\prod\limits_{\{i\}\in V}\mathbf{v}^{D}_{1,\sigma(\{i\})}\\
	&\mathop=\limits^{(a)}\prod\limits_{\{i\}\in V}(1-\mathbf{v}^{D}_{2,\sigma(\{i\})})
\end{align*}
(a) is based on the result of Lemma~\ref{lm:ncptvertex}. Since above result holds for $\forall V \in S, D\in \mathcal{D}$, the coordinates of any vertex in $H_d$ satisfy above relation. Then, let $\mathbf{x}_i=\mathbf{v}_{2,\sigma(\{i\})}$, we have
\begin{equation}
  \varepsilon'=\{\mathbf{x}\in\{0,1\}^n|\exists \mathbf{v}\in I_d\text{ s.t. }\mathbf{v}_{2,\sigma(\{i\})}=\mathbf{x}_i \}
\end{equation}

Based on the the results of Lemma~\ref{lm:cpt2org} and Lemma~\ref{lm:org2cpt}, there exists the one-one correspondence between the defender's pure strategy $D$ and $n$ coordinates of vertex $\mathbf{v}^D_{2, \sigma(\{i\})}, \forall 1\leq i\leq n$. Therefore, the set system $\varepsilon'$ is indeed the set system $\varepsilon$ defined in (\ref{def:setsystem}).

Therefore, we can write the defender oracle problem as the following $n-$dimensional pseudo-boolean function.
\begin{align*}
	&\max\limits_{\mathbf{v}\in I_d}\mathbf{w}^T\mathbf{v}=\max\limits_{\mathbf{v}\in I_d}\sum\limits_{V\in S}\mathbf{w}_{1,\sigma(V)}\mathbf{v}_{1,\sigma(V)}+\mathbf{w}_{2,\sigma(V)}\mathbf{v}_{2,\sigma(V)}\\
	&=\max\limits_{\mathbf{x}\in \varepsilon } \left[\sum\limits_{V\in S}\mathbf{w}_{1,\sigma(V)}\prod\limits_{\{i\}\in V} (1-\mathbf{x}_i)+\sum\limits_{V\in S}\mathbf{w}_{2,\sigma(V)}\prod\limits_{\{i\}\in V} \mathbf{x}_i\right]\\
	&=\max\limits_{\mathbf{x}\in \varepsilon } \left[\sum\limits_{V\in S}\mathbf{w}_{\sigma(V)}'\prod\limits_{\{i\}\in V} \mathbf{x}_i\right].
\end{align*}
Since the above procedure can be done in poly$(n)$ time, i.e., generate $\mathbf{w}_{\sigma(V)}'$ from $\mathbf{w}_{1,\sigma(V)}, \mathbf{w}_{2,\sigma(V)}$, the theorem follows.
\end{proof}

\end{document}